\documentclass[conference]{IEEEtran}

\usepackage{multirow}
\usepackage{siunitx}

\usepackage{amsmath,amssymb,amsfonts}
\usepackage{caption}
\usepackage{subcaption}
\usepackage{adjustbox}
\usepackage{colortbl}

\usepackage{listings}
\usepackage{url}
\usepackage{hhline} 

\usepackage{pifont}       
\usepackage{bbding}       
\usepackage{fontawesome}  
\usepackage{booktabs}
\usepackage{hyperref}
\usepackage{color}
\usepackage{xcolor}
\usepackage{pgfplots}
\pgfplotsset{compat=1.18}
\usepackage{algorithmic}
\usepackage{diagbox}

\usepackage{epstopdf}
\usepackage[most]{tcolorbox}  
\usepackage{lipsum}
\usepackage{listings}
\usepackage{tabularx}
\usepackage{float}
\usepackage[ruled,linesnumbered]{algorithm2e}
\usepackage{amsmath} 
\usepackage{multicol}
\usepackage{color}
\usepackage{xcolor}
\usepackage{xspace}
\usepackage{forest}
\usepackage{tikz}

\usepackage{comment}

\usepackage{enumitem}

\usepackage{enumitem}
\usepackage{graphicx}

\def\Snospace~{\S{}}

\definecolor{darkgreen}{rgb}{0.0, 0.5, 0.0} 
\newcommand{\cmark}{\textcolor{darkgreen}{\ding{51}}}%
\newcommand{\xmark}{\textcolor{red}{\ding{55}}}%

\definecolor{mGreen}{rgb}{0,0.6,0}
\definecolor{mGray}{rgb}{0.5,0.5,0.5}
\definecolor{mPurple}{rgb}{0.58,0,0.82}
\definecolor{backgroundColour}{rgb}{0.95,0.95,0.92}

\newcommand{\lstbg}[3][0pt]{{\fboxsep#1\colorbox{#2}{\strut #3}}}
\definecolor{codegreen}{rgb}{0,0.6,0}
\lstdefinelanguage{diff}{
	frame=single,
	basicstyle=\ttfamily\scriptsize\bfseries,
	morecomment=[f][\color{red}]{---}, 
	morecomment=[f][\color{codegreen}]{+++},
	morecomment=[f][\lstbg{red!20}]{-\ },
	morecomment=[f][\lstbg{green!20}]{+\ },
	morecomment=[f][\color{blue}]{@@},
}

\colorlet{punct}{red!60!black}
\definecolor{background}{HTML}{EEEEEE}
\definecolor{delim}{RGB}{20,105,176}
\colorlet{numb}{magenta!60!black}
\lstdefinelanguage{json}{
    basicstyle=\scriptsize\ttfamily,
    numbers=left,
    numberstyle=\scriptsize,
    stepnumber=1,
    numbersep=8pt,
    showstringspaces=false,
    breaklines=true,
    frame=lines,
    backgroundcolor=\color{background},
    literate=
      {:}{{{\color{punct}{:}}}}{1}
      {,}{{{\color{punct}{,}}}}{1}
      {\{}{{{\color{delim}{\{}}}}{1}
      {\}}{{{\color{delim}{\}}}}}{1}
      {[}{{{\color{delim}{[}}}}{1}
      {]}{{{\color{delim}{]}}}}{1},
}

\definecolor{codegreen}{rgb}{0,0.6,0}
\definecolor{codegray}{rgb}{0.5,0.5,0.5}
\definecolor{codepurple}{rgb}{0.58,0,0.82}
\definecolor{backcolour}{rgb}{0.95,0.95,0.88}

\definecolor{commentgreen}{HTML}{A3BE8C}    
\definecolor{bggray}{rgb}{0.95, 0.95, 0.95}
\definecolor{kwlavender}{rgb}{0.6, 0.4, 0.8}  
\definecolor{funcgreen}{rgb}{0.0, 0.5, 0.3}   
\definecolor{typecyan}{rgb}{0.0, 0.6, 0.8}    
\definecolor{linenumbergray}{rgb}{0.4, 0.4, 0.4}

\lstdefinelanguage{customC}{
  language=C++,
  morekeywords=[1]{if,else,for,while,return},
  morekeywords={[2]Modbus, UARTClass},
  morekeywords=[3]{uint32_t,int, uint8_t,char},
  sensitive=true,
  morecomment=[l]{//},
}

\definecolor{instr}{RGB}{38,38,150}
\definecolor{pseudo}{RGB}{38,140,96}
\definecolor{reg}{RGB}{175,175,125}

\lstdefinestyle{mystyle}{
    basicstyle=\ttfamily\footnotesize,
    frame=single,
    commentstyle=\color{codegreen},
    keywordstyle=\color{instr},
    numberstyle=\color{pseudo},
    stringstyle=\color{codepurple},
    breaklines=true,
    captionpos=b,
    keepspaces=true,
    numbers=left,
    numbersep=3pt,
    numberstyle=\ttfamily\footnotesize,
    showspaces=false,
    showstringspaces=false,
    showtabs=false，
    tabsize=2,
    framexleftmargin=0pt
}

\newcommand{\uemu}{{$\mu$Emu}\xspace}
\newcommand{\semu}{{SEmu}\xspace}

\newcommand{\fuzzware}{{Fuzzware}\xspace}
\newcommand{\gdma}{{GDMA}\xspace}
\newcommand{\hoedur}{{Hoedur}\xspace}

\newcommand{\otacap}{{OTACap}\xspace}
\newcommand{\firmline}{{FirmLine}\xspace}
\newcommand{\multifuzz}{MultiFuzz\xspace}
\newcommand{\icicle}{ICICLE\xspace}

\newcommand{\firmxray}{FirmXRay\xspace}
\newcommand{\aim}{{AIM}\xspace}
\newcommand{\aid}{{AidFuzzer}\xspace}
\newcommand{\rca}{{FirmRCA}\xspace}
\newcommand{\ppim}{{$P^{2}$IM}\xspace}
\renewcommand{\paragraph}[1]{\vspace{0.03in}\noindent\textbf{#1}}

\newcommand{\eg}{e.g.,}
\newcommand{\ie}{i.e.,}

\begin{document}
\title{SoK: A Large-Scale Empirical Study of Emulation-Based Dynamic Analysis Research for ARM Cortex-M Firmware}

\IEEEoverridecommandlockouts
\makeatletter\def\@IEEEpubidpullup{6.5\baselineskip}\makeatother
\IEEEpubid{\parbox{\columnwidth}{
		Network and Distributed System Security (NDSS) Symposium 2027\\
		22--26 March 2027, Seoul, Republic of Korea\\
		ISBN 978-1-970672-09-1\\  
		https://dx.doi.org/10.14722/ndss.2027.[23$|$24]xxxx\\
		www.ndss-symposium.org
}
\hspace{\columnsep}\makebox[\columnwidth]{}}

\author{
\IEEEauthorblockN{
Hongyuan Li$^{*}$,
Ke Wang$^{*}$,
Wei Zhou$^{*\text{\footnotesize\ding{41}}}$\thanks{Wei Zhou is the corresponding author.},
Le Guan$^{\dagger}$,
}

\IEEEauthorblockA{
$^{*}$School of Cyber Science and Engineering, Huazhong University of Science and Technology, China \\
$^{*}$Hubei Key Laboratory of Distributed System Security \\ 
$^{\dagger}$University of Georgia
}

\IEEEauthorblockA{
E-mails:
\{hornos2, ke\_wang\_tt, weizhou\_sec\}@hust.edu.cn,
leguan@uga.edu
}
}

\maketitle

\begin{abstract}

As microcontroller (MCU)-based devices become increasingly pervasive in daily
life, the need for efficient and scalable security analysis of MCU firmware
has become critical. While recent firmware re-hosting efforts have made
automated vulnerability assessment feasible, two major gaps remain
unaddressed. First, existing tools are commonly evaluated on limited and
heavily overlapping datasets, which undermines the generalizability of
reported results. Second, prior research typically advances the state of the
art along orthogonal dimensions, such as emulation, fuzzing, or bug diagnosis,
without providing a holistic understanding of how these components interact
and complement one another in dynamic analysis workflows.

Building on recently released large-scale MCU firmware datasets from \otacap
and \firmline, our work presents an empirical study of 24 emulation-based
firmware analysis tools published in top academic conferences and
journals. We systematically position these tools within a unified automated
analysis pipeline consisting of emulation configuration reconnaissance,
emulation, bug finding, and diagnosis. At each stage, we evaluate whether a
tool’s output provides sufficient information to enable the subsequent stage,
using a deduplicated and validated subset of 4,571 ARM Cortex-M based firmware samples.
Our results show that only 1,580 samples (34.5\%) can be successfully fuzzed, 
even when a sample is considered successful if at least one existing tool can fuzz it.
Among these, fuzzing results are generally poor,
with an average code coverage of only 10\% and a large number of
false crashes/hangs. Through a systematic analysis of failed fuzzing attempts and false-positive 
cases, we disclose fundamental challenges and methodological limitations in current approaches. 
These findings highlight critical gaps in the
state of the art and provide actionable insights to inform and guide future
research in emulation-based firmware analysis.

\end{abstract}

\IEEEpeerreviewmaketitle

\section{Introduction}

The number of connected IoT devices continues to grow rapidly~\cite{iotnumber2024}.
Microcontroller Units (MCUs) are widely adopted in IoT devices, including those deployed
in security-sensitive domains such as industrial control and healthcare systems,
owing to their low cost, low power consumption, and rich peripheral interfaces~\cite{nonlinux2024}.
However, despite their cost efficiency, MCUs lack many of the hardware-level security
mechanisms found in application processors, such as Intel SGX and even page-based
memory protection. Moreover, firmware is often implemented in memory-unsafe languages
such as C/C++, rendering it vulnerable to memory-related bugs.

As MCU-based devices continue to face active attacks, firmware security
testing has become increasingly critical. However, testing on real hardware
is often inefficient or even infeasible due to limited computational
resources, which significantly slow down the testing process. Furthermore,
many MCU devices are difficult or impossible to physically access once
deployed in environments such as industrial, aerospace, or healthcare
systems. As an alternative, firmware \textit{rehosting}---executing firmware on a host
with a different architecture and more powerful computing resources---has emerged as a practical and popular
testing strategy. Recent years have witnessed a surge of research focused on
developing more robust MCU firmware emulation environments~\cite{feng2020p2im,zhou2022your,tobias2022fuzzware,spensky2021conware,won2022what,zhou2021automatic,chong2024afriend,farrelly2023ember,wang2025aidfuzzer,scharnowski2025gdma},
which serve as the foundation for dynamic analysis techniques, particularly
automated vulnerability assessments such as fuzzing~\cite{scharnowski2023hoedur,chesser2024multifuzz,farrelly2023splits}. More
recently, researchers have also proposed dedicated bug diagnosis frameworks
tailored for rehosting-based firmware fuzzing~\cite{chang2025firmrca}.

\begin{figure*}[t]
\setlength\abovedisplayskip{0pt}
\setlength\belowdisplayskip{0pt}
\centering
\begin{adjustbox}{max width=\textwidth}
\includegraphics[width=\textwidth]{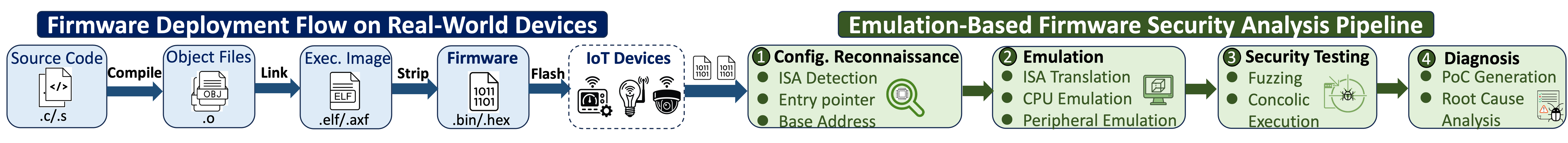}
\end{adjustbox}
\caption{Monolithic firmware development and emulation-based dynamic analysis pipeline.}
\label{fig:process}
\end{figure*}

This paper presents a Systematization of Knowledge (SoK) in the field of
emulation-based techniques for MCU firmware analysis, with a focus on the ARM Cortex-M architecture. Such a systematization is both timely and necessary for several reasons.

\textit{First}, the most relevant prior SoKs and
surveys~\cite{wright2021challenges,fasano2021sok} were published over five years
ago, when MCU-firmware rehosting was still at an early stage. Since then, more
than a dozen studies have advanced the state of the art across firmware
emulation, fuzzing, and downstream analysis. An updated SoK is therefore needed
to assess how much progress has been made, whether previously identified
challenges remain unresolved or have been mitigated, and what new challenges
have emerged.
\textit{Second}, existing surveys only cover limited topics in isolation, such as
emulation~\cite{fasano2021sok,zhou2025iot} and
fuzzing~\cite{asmita2025bare,qasem2021automatic,yun2022fuzzing}. However, a full
end-to-end firmware analysis pipeline involves at least four interdependent
stages, including firmware information recovery, emulation, fuzzing, and
root-cause analysis. By investigating them holistically within a unified
framework, we can reveal cross-stage dependencies and failure propagation that
remain hidden when studied in isolation.
\textit{Third}, prior evaluations have been constrained by the lack of a large-scale,
real-world MCU-firmware dataset. Existing techniques are commonly evaluated on
small and overlapping benchmarks, including SDK unit
tests~\cite{feng2020p2im,tobias2022fuzzware}, open-source demos for specific
development boards~\cite{zhou2022your,chong2024afriend}, and handcrafted
examples~\cite{gustafson2019toward,muench2018you}, making it difficult to assess
their generalizability and practical applicability. In particular, real-world
firmware can be more complex, heterogeneous and may lack the explicit
specifications assumed by existing methods. Prior
SoKs~\cite{wright2021challenges,fasano2021sok} therefore provide only 
qualitative summaries of high-level technical challenges,
existing mitigation strategies, and coarse-grained tradeoffs. Without really
evaluating them on a common, large-scale real-world benchmark, they cannot
quantitatively compare how much one approach improves emulation applicability or
vulnerability-discovery capability over another, nor can they explain the
underlying reasons for their failure cases with concrete evidence.

Recently, two large-scale MCU firmware datasets,
\otacap~\cite{nino2024unveiling} and \firmline~\cite{balgavy2024firmline}, have been introduced, collecting thousands of real-world MCU firmware samples from IoT devices’ companion mobile applications and other public sources. These datasets have thus far been leveraged for static analysis. We found them to be a
timely resource to enable scalable, data-driven studies of dynamic analysis
tools at unprecedented scale and diversity. Building on these resources, our SoK
conducts, to our knowledge, the first holistic empirical systematization of
emulation-based MCU-firmware analysis. We position 24 state-of-the-art tools
within a unified four-stage pipeline and evaluate the applicable tools on 4,571
unique ARM Cortex-M firmware samples selected from the two datasets. We focus on
the ARM Cortex-M architecture due to its prevalence in real IoT products (thus
the mentioned datasets) and first-class support in existing emulation-based
tools. Leveraging these datasets, our SoK highlights several evidence-based
findings. Through cross-stage validation and representative case studies, we
quantify the practical capabilities of current techniques, expose recurring
failure mechanisms and interoperability gaps, and explain why existing methods
struggle to scale to real-world firmware. In doing so, our SoK confirms,
refines, and deepens prior observations by showing how much progress has been
made, where and why current techniques still fail, and how future research can
address these limitations with evidence-based guidance.

\paragraph{Contributions.} This SoK advances the field of emulation-based
 firmware analysis techniques by:

\begin{itemize}[leftmargin=*]

    \item \textbf{Systematizing the knowledge.} We organize prior works within a
     unified analysis pipeline consisting of: (1) emulation configuration
     reconnaissance, (2) emulation, (3) security
     testing---particularly fuzzing, and (4) bug diagnosis. This 
     provides a unified perspective that contextualizes the relationships and
     interdependencies among existing techniques.

    \item \textbf{Revealing the gaps.} We conduct the first large-scale
     evaluation of SOTA MCU firmware re-hosting approaches using real-world
     datasets from \otacap~\cite{nino2024unveiling} and \firmline~\cite{balgavy2024firmline}, covering 4,571 ARM Cortex-M samples. Our analysis
     reveals that current tools achieve limited emulation success and
     coverage, generate numerous false crashes, and often misattribute root
     causes, underscoring the gap between academic prototypes and practical
     robustness.

    \item \textbf{Characterizing pitfalls and technical barriers.} We identify and analyze key challenges in large-scale evaluations at each workflow stage. These include difficulties in inferring emulation configurations for black-box firmware, incomplete ISA instruction and memory mapping support, inaccurate peripheral modeling, and various performance, implementation, and compatibility issues. Our insights are supported by real-world firmware examples and detailed false-crash artifacts.

    \item \textbf{Outlining future research directions.} We propose actionable
     avenues for advancing firmware re-hosting and dynamic analysis,
     including integrating hybrid analysis approaches, enhancing emulation
     fidelity, and developing more effective and scalable fuzzing and
     diagnostic strategies.

\end{itemize}

\section{Background}
\label{sec:background}

\subsection{MCU-Based Devices and Monolithic Firmware}

MCUs are widely used in deeply embedded systems. An MCU integrates a central
processing unit (CPU), memory, and various input/output peripherals within a single chip, in contrast to microprocessors used in personal computers or
other general-purpose applications, which typically consist of multiple
discrete components. MCUs employ diverse CPU architectures, including ARM,
MIPS, and RISC-V, with ARM being the most prevalent—powering over 60\% of
resource-constrained IoT devices~\cite{nonlinux2024}. A key characteristic
that distinguishes MCUs is their peripheral subsystem, which exhibits
substantial variation across different chip models and vendors.

The firmware executed on MCU-based embedded devices is typically monolithic,
encompassing application logic, device drivers, libraries, and, in some
cases, a real-time operating kernel (RTOS). MCU firmware is generally tailored to
specific tasks and often operates in an infinite loop, continuously receiving
input from the external environment through peripherals, processing the data,
and responding via actuators. Typically, the firmware interacts with
peripherals through three primary mechanisms: memory-mapped I/O
(MMIO) registers, interrupts, and direct memory access (DMA).

\subsection{Firmware Development and Deployment}
\label{sec:lifecycle}

Firmware is typically cross-compiled on a general-purpose computer to the
target instruction set of the MCU. The output of cross-compilation generally
follows standard executable formats (\eg~ELF or AXF), with additional
sections that may contain vendor-specific signatures or metadata. Before
deployment on real products, the ELF file is stripped of metadata, relocation
information, and symbols, resulting in a raw binary blob (\texttt{.bin}) or
an ASCII-encoded image (\texttt{.hex}) as demonstrated in~\autoref{fig:process}. In some cases, these files are
further compressed or encrypted before being programmed into the MCU’s
non-volatile memory. This build and deployment pipeline is not standardized,
allowing substantial flexibility for manufacturers to define their own
proprietary formats. Moreover, a firmware image may be partitioned into
multiple components, such as a bootloader, an application image, and a data
segment, which are typically flashed into separate memory regions. Many IoT
devices also support Over-the-Air (OTA) firmware updates, enabling
post-deployment maintenance, bug fixes, and feature enhancements.
Together, these factors make firmware emulation and analysis challenging.

\begin{figure*}[t]
\setlength\abovedisplayskip{0pt}
\setlength\belowdisplayskip{0pt}
\centering
\includegraphics[width=0.98\textwidth]{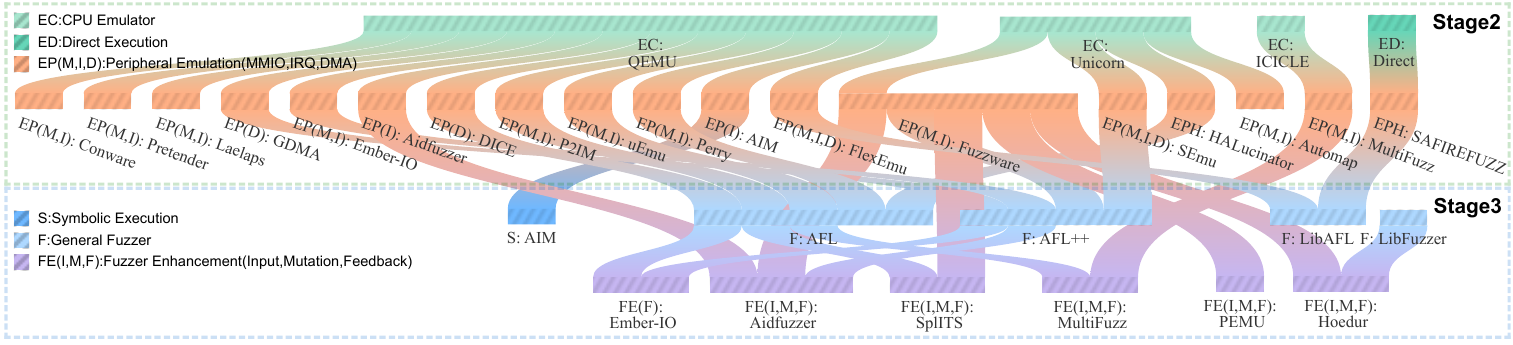}
\caption{Taxonomy and dependence relationships among emulation-based approaches in stage 2 and stage 3.}
\label{fig:summary}
\end{figure*}
\section{Systematization}

Firmware analysis introduces unique challenges not encountered in
general-purpose software analysis. Existing studies often tackle specific issues
within the broader firmware analysis pipeline. In this section, we first propose
a unified firmware analysis pipeline that incorporates the essential stages
involved in emulation-based firmware analysis. We then describe the literature
selection criteria in this SoK. Finally, we systematize existing research by
putting each work in the context of the proposed analysis pipeline.

\subsection{Emulation-Based Firmware Analysis Pipeline}

As illustrated in~\autoref{fig:process}, 
assuming a firmware image is available, 
four main stages are involved in emulation-based firmware analysis.

\paragraph{Stage 1: Emulation Configuration Reconnaissance.} As discussed
 in~\autoref{sec:lifecycle}, real-world firmware often lacks symbols and
 memory layout information.  Moreover, firmware is tightly coupled with its
 hardware platform.  To execute such firmware within an emulator, it is
 essential to understand the underlying hardware characteristics, including
 the instruction set architecture (ISA), the memory layout (at least the base address), 
 and the entry point.  We refer to this information collection process
 as the \textit{emulation configuration},
 which is fundamental to correctly initializing a working emulation environment.

\paragraph{Stage 2: Emulation.} With the emulation configuration established,
 the next stage aims to achieve full-system emulation, enabling firmware to
 execute in a completely re-hosted environment on a general-purpose
 CPU without relying on physical hardware. This stage primarily involves ISA
 translation and peripheral interaction emulation. The latter imposes a huge
 challenge and is a main focus of existing works.

\paragraph{Stage 3: Security Testing.} Once the firmware can execute reliably
 within the emulator, dynamic security analysis tools---particularly
 fuzzers---can be integrated to uncover vulnerabilities. In this stage,
 (1) fuzzing inputs are supplied when the firmware reads from peripheral
 registers or DMA buffers, (2) code coverage is monitored through dynamic
 instrumentation, and (3) abnormal or crash behaviors are detected by
 observing invalid memory accesses and other runtime faults.

\paragraph{Stage 4: Diagnosis.} After a crash is detected, bug diagnosis
 becomes a necessary step toward resolving the issue. This stage involves
 identifying the root cause, confirming the exploitability and the impact, and
 generating proof-of-concept evidence for validation.

\begin{table*}[t]
\centering
\caption{Dependence information and summary of emulation-based approaches in stage 2/3 ($\varnothing$: no dependence/enhancement)}
\begin{adjustbox}{max width=\textwidth}
\begin{tabular}{|l|c|c|cccc|c|ccc|cc|}
\hline
\multicolumn{1}{|c|}{\multirow{4}{*}{\textbf{Tool [reference]}}} & \multirow{4}{*}{\textbf{Venue}} & \multirow{4}{*}{\textbf{\rotatebox{90}{Avail.}}} & \multicolumn{4}{c|}{\multirow{3}{*}{\textbf{\begin{tabular}[c]{@{}c@{}}Dependence \\ Information\\ In Stage 1\end{tabular}}}} & \multirow{3}{*}{\textbf{\begin{tabular}[c]{@{}c@{}}EC/ET: \\ CPU Emulation/\\ Transplanted\end{tabular}}} & \multicolumn{3}{c|}{\multirow{3}{*}{\textbf{\begin{tabular}[c]{@{}c@{}}EP: Peripheral Emulation/\\ EPH: HAL Replacement\end{tabular}}}} & \multicolumn{2}{c|}{\multirow{3}{*}{\textbf{\begin{tabular}[c]{@{}c@{}}Security \\ Testing\end{tabular}}}} \\
\multicolumn{1}{|c|}{}                                          &                                 &                                                       & \multicolumn{4}{c|}{}                                                                                                       &                                                                                                           & \multicolumn{3}{c|}{}                                                                                                                   & \multicolumn{2}{c|}{}                                                                                      \\
\multicolumn{1}{|c|}{}                                          &                                 &                                                       & \multicolumn{4}{c|}{}                                                                                                       &                                                                                                           & \multicolumn{3}{c|}{}                                                                                                                   & \multicolumn{2}{c|}{}                                                                                      \\ \cline{4-13} 
\multicolumn{1}{|c|}{}                                          &                                 &                                                       & \multicolumn{1}{c|}{\textbf{ISA}}  & \multicolumn{1}{c|}{\textbf{Base}} & \multicolumn{1}{c|}{\textbf{EP}} & \textbf{Other} & \textbf{Tested ISA}                                                                                       & \multicolumn{1}{c|}{\textbf{MMIO}}                   & \multicolumn{1}{c|}{\textbf{IRQ}}                  & \textbf{DMA}                & \multicolumn{1}{c|}{\textbf{Method}}                            & \textbf{Enhancement}                             \\ \hline \hline
Pretender~\cite{gustafson2019toward}                            & RAID'19                         & \cmark                                                & \multicolumn{1}{c|}{\checkmark}    & \multicolumn{1}{c|}{\checkmark}    & \multicolumn{1}{c|}{\checkmark}  & HL             & EC:ARM-M                                                                                                  & \multicolumn{2}{c|}{HL-Base}                                                                             & \xmark                       & \multicolumn{1}{c|}{F:Fuzz}              & FE($\varnothing$,$\varnothing$,$\varnothing$,$\varnothing$)     \\ \hline
$P^{2}$IM~\cite{feng2020p2im}                                   & SEC'20                            & \cmark                                                & \multicolumn{1}{c|}{\checkmark}    & \multicolumn{1}{c|}{\checkmark}    & \multicolumn{1}{c|}{\checkmark}  & $\varnothing$  & EC:ARM-M                                                                                                  & \multicolumn{1}{c|}{Pattern-base}                  & \multicolumn{1}{c|}{RR}                             & \xmark                       & \multicolumn{1}{c|}{F:Fuzz}              & FE($\varnothing$,$\varnothing$,$\varnothing$,$\varnothing$)     \\ \hline
HALucinator~\cite{clements2020halucinator}                      & SEC'21                            & \cmark                                                & \multicolumn{1}{c|}{\checkmark}    & \multicolumn{1}{c|}{\checkmark}    & \multicolumn{1}{c|}{\checkmark}  & HC             & EC:ARM-M                                                                                                  & \multicolumn{3}{c|}{EPH:HAL Replacement}                                                                                                & \multicolumn{1}{c|}{F:Fuzz}              & FE($\varnothing$,$\varnothing$,$\varnothing$,$\varnothing$)     \\ \hline
Laelaps~\cite{cao2020device}                                    & ACSAC'21                          & \cmark                                                & \multicolumn{1}{c|}{\checkmark}    & \multicolumn{1}{c|}{\checkmark}    & \multicolumn{1}{c|}{\checkmark}  & $\varnothing$  & EC:ARM-M                                                                                                  & \multicolumn{1}{c|}{PC-base}                       & \multicolumn{1}{c|}{Random}                         & \xmark                       & \multicolumn{1}{c|}{Unsupported}              & -     \\ \hline
$\mu$Emu~\cite{zhou2021automatic}                               & SEC'21                          & \cmark                                                & \multicolumn{1}{c|}{\checkmark}    & \multicolumn{1}{c|}{\checkmark}    & \multicolumn{1}{c|}{\checkmark}  & $\varnothing$  & EC:ARM-M                                                                                                  & \multicolumn{1}{c|}{PC-base}                       & \multicolumn{1}{c|}{RR}                             & \xmark                       & \multicolumn{1}{c|}{F:Fuzz}              & FE($\varnothing$,$\varnothing$,$\varnothing$,$\varnothing$)     \\ \hline
Conware~\cite{spensky2021conware}                               & Asia CCS'21                     & \cmark                                                & \multicolumn{1}{c|}{\checkmark}    & \multicolumn{1}{c|}{\checkmark}    & \multicolumn{1}{c|}{\checkmark}  & HL             & EC:ARM-M                                                                                                  & \multicolumn{2}{c|}{HL-Base}                                                                             & \xmark                       & \multicolumn{1}{c|}{Unsupported}              & -     \\ \hline
DICE~\cite{mera2021dice}                                        & S\&P'21                         & \cmark                                                & \multicolumn{1}{c|}{\checkmark}    & \multicolumn{1}{c|}{\checkmark}    & \multicolumn{1}{c|}{\checkmark}  & $\varnothing$  & EC:ARM-M,MIPS32                                                                                           & \multicolumn{2}{c|}{P2IM}                                                                                & Pattern-base                 & \multicolumn{1}{c|}{F:Fuzz}              & FE($\varnothing$,$\varnothing$,$\varnothing$,$\varnothing$)     \\ \hline
Fuzzware~\cite{tobias2022fuzzware}                              & SEC'22                          & \cmark                                                & \multicolumn{1}{c|}{\checkmark}    & \multicolumn{1}{c|}{\checkmark}    & \multicolumn{1}{c|}{\checkmark}  & $\varnothing$  & EC:ARM-M                                                                                                  & \multicolumn{1}{c|}{PC-base}                       & \multicolumn{1}{c|}{RR,Fuzz}                        & \xmark                       & \multicolumn{1}{c|}{F:Fuzz}              & FE($\varnothing$,$\varnothing$,$\varnothing$,$\varnothing$)     \\ \hline
SEmu~\cite{zhou2022your}                                        & CCS'22                          & \cmark                                                & \multicolumn{1}{c|}{\checkmark}    & \multicolumn{1}{c|}{\checkmark}    & \multicolumn{1}{c|}{\checkmark}  & Spec.          & EC:ARM-M                                                                                                  & \multicolumn{2}{c|}{Spec.-base}                                                                          & Manual                       & \multicolumn{1}{c|}{F:Fuzz}              & FE($\varnothing$,$\varnothing$,$\varnothing$,$\varnothing$)     \\ \hline
AIM~\cite{feng2023aim}                                          & TDSC'23                         & \cmark                                                & \multicolumn{1}{c|}{\checkmark}    & \multicolumn{1}{c|}{\checkmark}    & \multicolumn{1}{c|}{\checkmark}  & $\varnothing$  & EC:ARM-M                                                                                                  & \multicolumn{1}{c|}{P2IM}                          & \multicolumn{1}{c|}{WS-base}                 & \xmark                       & \multicolumn{1}{c|}{S:Symbolic}          & -                                                               \\ \hline
Ember-IO~\cite{farrelly2023ember}                               & Asia'23                         & \cmark                                                & \multicolumn{1}{c|}{\checkmark}    & \multicolumn{1}{c|}{\checkmark}    & \multicolumn{1}{c|}{\checkmark}  & MM             & EC:ARM-M                                                                                                  & \multicolumn{1}{c|}{IP-base}                       & \multicolumn{1}{c|}{RR}                             & \xmark                       & \multicolumn{1}{c|}{F:Fuzz}              & FE($\varnothing$,$\varnothing$,F,$\varnothing$)                 \\ \hline
ICICLE~\cite{chesser2023icicle}                                 & ISSTA'23                        & \cmark                                                & \multicolumn{1}{c|}{\checkmark}    & \multicolumn{1}{c|}{\checkmark}    & \multicolumn{1}{c|}{\checkmark}  & $\varnothing$  & EC:ARM-M,MSP430                                                                                              & \multicolumn{1}{c|}{Fuzzware}                      & \multicolumn{1}{c|}{RR}                             & \xmark                       & \multicolumn{1}{c|}{F:Fuzz}              & FE(I,M,$\varnothing$,$\varnothing$)                             \\ \hline
Hoedur~\cite{scharnowski2023hoedur}                             & SEC'23                          & \cmark                                                & \multicolumn{1}{c|}{\checkmark}    & \multicolumn{1}{c|}{\checkmark}    & \multicolumn{1}{c|}{\checkmark}  & $\varnothing$  & EC:ARM-M                                                                                                  & \multicolumn{1}{c|}{Fuzzware}                      & \multicolumn{1}{c|}{RR,Fuzz}                        & \xmark                       & \multicolumn{1}{c|}{F:Fuzz}              & FE(I,M,F,$\varnothing$)                                         \\ \hline
SplITS~\cite{farrelly2023splits}                                & ESORICS'23                      & \cmark                                                & \multicolumn{1}{c|}{\checkmark}    & \multicolumn{1}{c|}{\checkmark}    & \multicolumn{1}{c|}{\checkmark}  & SC             & EC:ARM-M                                                                                                  & \multicolumn{1}{c|}{Fuzzware}                      & \multicolumn{1}{c|}{RR}                             & \xmark                       & \multicolumn{1}{c|}{F:Fuzz}              & FE(I,M,F,$\varnothing$)                                         \\ \hline
SAFIREFUZZ~\cite{2023_safirefuzz}                               & SEC'23                          & \cmark                                                & \multicolumn{1}{c|}{\checkmark}    & \multicolumn{1}{c|}{\checkmark}    & \multicolumn{1}{c|}{\checkmark}  & HC             & ET:ARM-M                                                                                                  & \multicolumn{3}{c|}{EPH:HAL Replacement}                                                                                                & \multicolumn{1}{c|}{F:Fuzz}              & FE($\varnothing$,$\varnothing$,F)                               \\ \hline
Perry~\cite{chong2024afriend}                                   & SEC'24                          & \cmark                                                & \multicolumn{1}{c|}{\checkmark}    & \multicolumn{1}{c|}{\checkmark}    & \multicolumn{1}{c|}{\checkmark}  & HC             & EC:ARM-M                                                                                                  & \multicolumn{2}{c|}{HC-base}                                                                             & \xmark                       & \multicolumn{1}{c|}{F:Fuzz}              & FE($\varnothing$,$\varnothing$,$\varnothing$,$\varnothing$)     \\ \hline
MultiFuzz~\cite{chesser2024multifuzz}                           & SEC'24                          & \cmark                                                & \multicolumn{1}{c|}{\checkmark}    & \multicolumn{1}{c|}{\checkmark}    & \multicolumn{1}{c|}{\checkmark}  & $\varnothing$  & EC:ARM-M                                                                                                  & \multicolumn{1}{c|}{PC-Base}                       & \multicolumn{1}{c|}{RR,Fuzz}                        & \xmark                       & \multicolumn{1}{c|}{F:Fuzz}              & FE(I,M,F,$\varnothing$)                                         \\ \hline
GDMA~\cite{scharnowski2025gdma}                                 & SEC'25                          & \cmark                                                & \multicolumn{1}{c|}{\checkmark}    & \multicolumn{1}{c|}{\checkmark}    & \multicolumn{1}{c|}{\checkmark}  & $\varnothing$  & EC:ARM-M                                                                                                  & \multicolumn{2}{c|}{Fuzzware}                                                                            & Pattern-base                 & \multicolumn{1}{c|}{F:Fuzz}              & FE($\varnothing$,$\varnothing$,$\varnothing$,$\varnothing$)     \\ \hline
Aidfuzzer~\cite{wang2025aidfuzzer}                              & SEC'25                          & \cmark                                                & \multicolumn{1}{c|}{\checkmark}    & \multicolumn{1}{c|}{\checkmark}    & \multicolumn{1}{c|}{\checkmark}  & $\varnothing$  & EC:ARM-M                                                                                                  & \multicolumn{1}{c|}{Fuzzware}                      & \multicolumn{1}{c|}{WS-base}                        & \xmark                       & \multicolumn{1}{c|}{F:Fuzz}              & FE (I,M,F,$\varnothing$)                                        \\ \hline
FlexEmu~\cite{lei2025flexemu}                                   & CCS'25                          & \xmark                                                & \multicolumn{1}{c|}{\checkmark}    & \multicolumn{1}{c|}{\checkmark}    & \multicolumn{1}{c|}{\checkmark}  & HC             & EC:ARM-M                                                                                                  & \multicolumn{3}{c|}{LLM-assisted HC-base}                                                                                               & \multicolumn{1}{c|}{F:Fuzz}              & FE($\varnothing$,$\varnothing$,$\varnothing$,$\varnothing$)     \\ \hline
PEmu~\cite{bley2025protocol}                                    & CCS'25                          & \cmark                                                & \multicolumn{1}{c|}{\checkmark}    & \multicolumn{1}{c|}{\checkmark}    & \multicolumn{1}{c|}{\checkmark}  & $\varnothing$  & EC:ARM-M                                                                                                  & \multicolumn{2}{c|}{Fuzzware/SEmu/Hoedur}                                                                & \xmark                       & \multicolumn{1}{c|}{F:Fuzz}              & FE (I,M,F,$\varnothing$)                                        \\ \hline
\end{tabular}
\end{adjustbox}
\label{tab:depend}
\flushleft
\scriptsize{
\textbf{Dependence Information:} ISA, Base Address, and Entry Point are required by all tools. \\
\textbf{Notations:} 
ARM-M = ARM Cortex-M HC = HAL Code, MM = ELF Format Memory Map
HL= Hardware-log, 
WS = Waiting State,
PC = Path Constraints, 
IP=Input Playback.\\
\textbf{Enhancement, FE (I,M,F,C):} I = Input Generation, M = Input Mutation, F = Execution Feedback, C = Crash Detection.
}
\end{table*}

\subsection{Literature Scope and Selection Criteria}

We initiated our exploration of emulation-based firmware analysis research by
examining publications from four prominent security conferences: IEEE S\&P,
USENIX Security, ACM CCS, and NDSS, spanning the past 15 years (2010–2025). We
focused on papers containing the keywords \textit{MCU}, \textit{Monolithic},
\textit{firmware}, \textit{re-hosting} and \textit{emulation} for at least two
occurrences. Additionally, we included cited works from other reputable security
and software engineering venues, as detailed in Appendix~\ref{app:litscope}. A
candidate needs to further meet the following criteria to be included in our
study. We note that such criteria may exclude some works that are relevant to
firmware emulation. We discuss how this decision influences the findings of this SoK
in~\autoref{sec:limited-finding-scope}. 

\begin{itemize}

\item \textbf{Monolithic Firmware.} We only target monolithic firmware,
corresponding to Type-II and Type-III firmware as defined
in~\cite{muench2018you}. Linux-based firmware analysis research (Type-I)
(\eg~\cite{chen2016towards,zheng2019firm,kim2020firmae,
tay2023greenhouse,xiao2025housefuzz,costin2014large,kammerstetter2014prospect})
which presents very different characteristics and challenges is excluded.

\item \textbf{Life-Cycle Emulation.} The work must emulate the full life-cycle
of firmware execution. Therefore, hardware-in-the-loop (HITL) approaches
(\eg~\cite{zaddach2014avatar,mera2024shift,li2022muafl, won2022what,liu2024co3})
and approaches that emulate only a partial execution phase, such as startup-only
emulation (\eg~Jetset~\cite{johnson2021jetset}), are excluded. This criterion
ensures objective evaluation of the work's genuine emulation capability. 

\item \textbf{Generality.} The work should support general firmware analysis
rather than being specialized for a particular platform or domain that requires
substantial manual adaptation. Analysis of baseband
firmware~\cite{hernandez2022firmwire,kim2021basespec,klischies2025basebridge},
automotive systems~\cite{chen2022metaemu}, industrial-control
systems~\cite{clements2021your}, satellite firmware~\cite{scharnowski2023case},
USB firmware~\cite{peng2020usbfuzz,hernandez2017firmusb}, and TrustZone
software~\cite{harrison2020partemu} are examples of such work.

\item \textbf{ARM Cortex-M.} While MCUs encompass a wide range of architectures,
ARM Cortex-M is the dominant one. Therefore, general firmware emulation tools
treat ARM Cortex-M as a first-class target. We exclude tools that do not
support ARM Cortex-M, including those specialized for
MSP430~\cite{davidson2013fie} and x86~\cite{yin2023rsfuzzer}. This eliminates
architecture-specific bias, allowing for a more meaningful comparison of the
emulation technique itself.

\item \textbf{Dynamic Analysis.} The emulation technique must support at least
one dynamic analysis task, particularly fuzzing. We exclude partial-emulation
approaches designed primarily to assist static analysis
(\eg~\cite{yao2019identifying,tsang2024ffxe}).

\end{itemize}

In the end, 21 works in stage 2/3 were selected, as summarized
in~\autoref{tab:depend}. \autoref{fig:summary} visualizes their technological
interdependency. For example, most emulators are based on QEMU. For stage 1, we
included \otacap~\cite{nino2024unveiling} and
\firmline~\cite{balgavy2024firmline} due to their contributions in firmware
recognition. For stage 4, FirmRCA~\cite{chang2025firmrca} is the only
post-fuzzing tool that we found in the literature. 

\subsection{Stage 1: Emulation Configuration Reconnaissance}

Current firmware re-hosting techniques require detailed knowledge of the
MCU model, specifically its \textbf{ISA}, base address
(\textbf{Base}), and entry point (\textbf{EP}). 
Some tools also require other
resources such as MCU specifications~\cite{zhou2022your}, I/O operation logs
from the original hardware~\cite{spensky2021conware,gustafson2019toward}, or
peripheral hardware abstraction layer (HAL) code or complete source code~\cite{clements2020halucinator,chong2024afriend,farrelly2023splits}. 
We list such dependency information for each tool in the middle columns of~\autoref{tab:depend}. 

However, as mentioned in \autoref{sec:lifecycle}, these assumptions often do not apply to real-world firmware analysis tasks, as real-world firmware is typically a stripped black-box binary blob. Therefore, inferring the aforementioned information becomes a
prerequisite for automated firmware emulation and analysis.
Currently, only a few studies~\cite{wen2020firmxray,nino2024unveiling,balgavy2024firmline} explicitly address
these challenges (ISA, Base, EP), all relying on heuristic-based approaches.

\firmline supports \textbf{ISA} detection with a statistical method that
 analyzes binary features to infer possible architectures and validates the
 results through reverse engineering, considering the architecture correct if
 valid function boundaries can be identified. For \textbf{Base} detection, \firmxray~\cite{wen2020firmxray} identifies potential
 absolute addresses, such as (1) function pointers used for indirect
 branches,
 (2) pointers to strings passed as function arguments, and
 (3) interrupt vector table (IVT) entries.  
 It then selects a candidate base address that resolves the largest number of pointers correctly.  
\otacap~\cite{nino2024unveiling} extends \firmxray by generalizing base
 address detection beyond specific vendors such as TI and Nordic, while
 maintaining similar constraints on indirect calls, string pointers, and
 vector table entries. Likewise, \firmline relies primarily on absolute
 function pointers for base address resolution. For \textbf{EP} detection, \firmxray assumes that the IVT is always located at the beginning of the firmware and extracts the reset handler address from it.

\subsection{Stage 2: Emulation}

Following system-on-chip (SoC) design principles, an MCU integrates multiple
hardware components.  We classify existing emulation-based studies according to
the primary component they target:  CPU emulation (\textbf{EC}), peripheral
emulation (\textbf {EP}),  peripheral hardware abstraction layer (HAL)
replacement (\textbf {EPH}),  and transplanted execution (\textbf{ET}).

\subsubsection{\textbf{EC:} CPU Emulation}

MCU devices exhibit significant diversity in CPU architectures, with ARM, MIPS,
and RISC-V dominating the market. Tools such as QEMU~\cite{bellard2005qemu} and
Unicorn~\cite{quynh2015unicorn} already support a broad range of widely used
embedded architectures. While QEMU is more comprehensive, Unicorn is lightweight
and optimized for dynamic binary analysis at the cost of sacrificing some
feature support, such as multimedia instructions. In addition,
\icicle~\cite{chesser2023icicle} leverages the SLEIGH specifications from the
Ghidra framework~\cite{ghidra} to achieve CPU emulation, enabling flexible and
efficient dynamic binary instrumentation. These tools serve as the foundation
for many existing works, as illustrated in~\autoref{fig:summary}.  

\subsubsection{\textbf{EP:} Peripheral Emulation}

As noted in~\autoref{sec:background}, there are three primary types of
peripheral I/O: MMIO, IRQ, and DMA. Accurate modeling of these behaviors is essential for firmware to interact correctly with its external environment, which is crucial for meaningful dynamic analysis. Peripheral emulation addresses this need by modeling these peripheral I/O operations, allowing firmware to execute as if it were on actual hardware.

\paragraph{Heuristic-Based Peripheral Modeling.} To emulate \textit{peripheral MMIO}
 operations, most existing solutions adopt different heuristic-based methods to infer
 peripheral behavior from firmware execution behaviors, as summarized in~\autoref{tab:depend}. 
P$^{2}$IM~\cite{feng2020p2im} first inferred register types from MMIO access
patterns and generated corresponding read responses, but its assumptions are
limited for complex peripherals. Subsequent work
~\cite{johnson2021jetset,cao2020device,tobias2022fuzzware,zhou2021automatic}
incorporates path constraints, assuming that invalid peripheral responses lead
to abnormal execution. $\mu$Emu~\cite{zhou2021automatic} uses symbolic
execution to learn feasible peripheral responses, whereas
Fuzzware~\cite{tobias2022fuzzware} relaxes such constraints and exposes
peripheral reads to fuzzing inputs for broader exploration. Ember-IO
~\cite{farrelly2023ember} further reuses coverage-increasing fuzzing values as
peripheral responses. Although effective for exploration, such heuristic
modeling may sacrifice fidelity and introduce false positives~\cite{zhou2025iot}.

For \textit{interrupt timing modeling}, as summarized in~\autoref{tab:depend}, current solutions typically adopt ad-hoc strategies such
as the round-robin (\textbf{RR}) method by default. 
In this approach, an enabled peripheral
interrupt is triggered after every fixed number of basic blocks (\eg~1,000).  When multiple peripheral interrupts are enabled, the triggers
are cycled in sequence.  
Tools such as \fuzzware and \multifuzz
further support a fuzzing-guided interrupt triggering method (\textbf{Fuzz}), 
where the interrupt intervals are determined dynamically based on fuzzing inputs.  
AIM and \aid have recently noted that fixed or fuzzing-guided strategies can cause firmware to remain in prolonged waiting states. To address this, they propose a context-aware approach that triggers interrupts when the firmware encounters \textbf{waiting state} conditions. These conditions include executing sleep instructions, accessing global objects that can be modified by interrupt service routines (ISRs), or looping indefinitely during configuration phases.

For \textit{DMA modeling}, DICE~\cite{mera2021dice} infers DMA behavior from
heuristic MMIO access patterns but assumes relatively fixed register layouts.
GDMA~\cite{scharnowski2025gdma} instead iteratively learns DMA models from
memory-access traces, supporting more diverse DMA behaviors.

\paragraph{High-Fidelity Modeling.} Inferring peripheral behavior solely from firmware execution is incomplete and inaccurate compared to real hardware. Researchers strive for higher-fidelity emulation by incorporating more detailed hardware knowledge. Pretender~\cite{gustafson2019toward} and Conware~\cite{spensky2021conware} use peripheral I/O logs from real hardware or heuristic algorithms to develop peripheral models. However, this approach still depends on physical devices and may not fully capture hardware behaviors due to the limited scope and duration of the logs collected. To overcome hardware dependence, SEmu~\cite{zhou2022your} applies natural language processing (NLP) to extract peripheral behavior specifications from hardware manuals, translating them into executable models. Similarly, Perry~\cite{chong2024afriend} creates high-fidelity peripheral emulator models by learning from open-source peripheral HAL code using symbolic execution. FlexEmu~\cite{lei2025flexemu} further advances this by employing an LLM to automatically construct flexible and accurate peripheral models from open-source peripheral HAL code.

\subsubsection{\textbf{EPH:} HAL Replacement} 

Instead of directly modeling peripheral behavior, the hardware abstraction
layer (HAL) of the firmware  can be replaced with host-side
implementations~\cite{clements2020halucinator} to circumvent peripheral
emulation challenges.  These approaches rely on identifying HAL functions
from open-source driver code.  However, not all firmware employs an
HAL layer, and the overall effectiveness of this strategy  largely depends on
the accuracy of library matching algorithms used in HAL detection.

\subsubsection{\textbf{ET:} Transplanted Execution}

Since ISA translation is computationally expensive, several studies have
proposed transplanting embedded programs  to enable execution on high-end CPUs.  
For example, SAFIREFUZZ~\cite{2023_safirefuzz} allows ARM Cortex-M
firmware to execute on ARM Cortex-A servers  without CPU emulation by
employing HAL function replacement and dynamic binary
rewriting  to handle peripheral interactions and accommodate
Cortex-M–specific instructions.

\subsection{Stage 3: Security Testing}
\label{sec:fuzzapp}

This section discusses security testing techniques enabled by firmware
emulation.

\subsubsection{\textbf{F:} Fuzzing}

Firmware emulation is primarily employed to enable fuzzing-based firmware testing.  
This is because fuzzing, as a randomized testing technique, does not require highly accurate emulation  
but instead focuses on maximizing code coverage.  
Most existing emulators integrate general-purpose fuzzers such as  
AFL/AFL++~\cite{AFLplusplus-Woot20}, LibAFL~\cite{libafl}, and libFuzzer~\cite{libfuzzer}  
as front-ends for input generation and mutation, as illustrated in~\autoref{fig:summary}.  
The emulation environment serves as the back-end, responsible for test case execution,  
feedback collection, and crash detection.

\begin{table}[t]
\centering
\caption{Fuzzing enhancement methods (—: indicates no enhancement with its integrated general fuzzer in~\autoref{fig:summary})
When only one element is present in the cell and preceded by a ``+'', it signifies the addition of new feedback to the original edge coverage.)
}
\label{tab:fe}
\begin{adjustbox}{width=\columnwidth}
\begin{tabular}{|l|c|c|}
\hline
\multicolumn{1}{|c|}{\textbf{Tool}} & \textbf{Input Generation\&Mutation} & \textbf{Feedback}      \\ \hline\hline
Ember-IO                          & —                                & FERMCov                \\ \hline
ICICLE                            & Cmplog                              & +CompareCov             \\ \hline
Hoedur                            & Input Extension+MultiStream         & FERMCov                \\ \hline
SplITS                            & I2S                                 & FERMCov+CompareCov            \\ \hline
SAFIREFUZZ                        & —                                & Non-colliding Coverage \\ \hline
MultiFuzz                         & Input Extension+MultiStream+I2S     & BB Coverage            \\ \hline
Aidfuzzer                         & Input Extension+MultiStream         & FERMCov                \\ \hline
PEMU & Protocol-aware & +Protocol-specific Probes \\ \hline
\end{tabular}
\end{adjustbox}
\end{table}

\subsubsection{\textbf{FE:} Enhancement}
\label{sec:fe}

Recent studies have shown that general-purpose fuzzers struggle to handle firmware-specific behaviors,  
resulting in reduced effectiveness and efficiency.  
Existing work therefore enhances fuzzing mainly through
execution feedback and input generation/mutation, as summarized
in~\autoref{tab:fe}.

For \textbf{execution feedback}, Ember-IO~\cite{farrelly2023ember} introduces
``FERMCov'' to distinguish ISR-induced edges from normal execution and reduce
coverage nondeterminism. ICICLE~\cite{chesser2023icicle} and
MultiFuzz~\cite{chesser2024multifuzz} further use P-code instrumentation to
provide finer-grained branch-level feedback.

For \textbf{input generation and mutation}, Hoedur~\cite{scharnowski2023hoedur}
and MultiFuzz model firmware inputs as multiple peripheral streams rather than a
single linear stream, enabling stream-aware scheduling and mutation.
MultiFuzz additionally adopts ICICLE instrumentation for ``Input-to-State (I2S)''
replacement~\cite{aschermann2019redqueen}, while PEmu~\cite{bley2025protocol}
introduces protocol-aware fuzzing for protocol-oriented firmware.

\begin{table}[t]
    \centering
    \caption{The Generic Memory Layout and Permission Configuration in \fuzzware/\hoedur/\multifuzz}
    \label{tab:memorymap}
    \begin{adjustbox}{width=\columnwidth}
    \begin{tabular}{|l|r|r|c|}
         \hline
         Segment & \multicolumn{1}{c|}{\textbf{Base}} & \multicolumn{1}{c|}{\textbf{Size}} & \textbf{Permission} \\
         \hline\hline
         Text & User-defined & File size & r-x \\
         \hline
         Ram & \verb|0x2000_0000| & \verb|0x10_0000| & rw- \\
         \hline
         Peripheral & \verb|0x4000_0000| & \verb|0x2000_0000| & rw- \\
         \hline
         System Peripheral & \verb|0xE000_0000| & \verb|0x1000_5000| & rw- \\
         \hline
         IRQ\_RET & \verb|0xFFFF_F000| & \verb|0x1000| & --x \\
         \hline
    \end{tabular}
    \end{adjustbox}
\end{table}

For \textbf{crash/hang detection}, firmware fuzzers typically lack sanitizers
and instead rely on coarse memory maps, access permissions, and emulator
exceptions (see~\autoref{tab:memorymap}). We normalize their failure reports
into four categories: \textit{Unmapped Access}, including read (\textbf{UR}),
write (\textbf{UW}), and instruction fetch (\textbf{UF}) outside mapped regions;
\textit{Permission Violation}, including invalid read (\textbf{RV}), write
(\textbf{WV}), and execution (\textbf{EV}); \textit{Unsupported Instruction}
(\textbf{UI}); and \textit{Hang}, where execution fails to consume input within
a time or execution-count threshold (\eg~three million basic blocks in Hoedur/MultiFuzz). Appendix~\ref{app:crash/hang-map} maps tool-specific
exceptions to these categories.

\subsubsection{\textbf{S:} Symbolic Execution}
Beyond fuzzing, emulation has also been employed to facilitate symbolic execution.  
For example, \aim~\cite{feng2023aim} leverages symbolic execution to predict interrupt-triggering behavior. 
Rather than integrating directly with fuzzing, \aim\ symbolizes data register reads and uses concolic execution for crash detection.
Nonetheless, it depends on a basic memory error detector that aligns with the memory layout and permissions detailed in~\autoref{tab:memorymap}.

\subsection{Stage 4: Diagnosis}

Bug diagnosis aims to identify the root cause of observed crashes  
and represents a critical yet time-consuming postmortem analysis task.  
MCU firmware, typically monolithic, lacks common fault detection mechanisms  
such as segmentation faults, leading to undetected memory corruptions—often referred to as ``silent'' bugs  
in prior studies~\cite{muench2018you}.  
As a result, crashes in MCU firmware often occur ``late'' long after the initial memory corruption,  
which significantly complicates bug diagnosis and increases the manual effort required.  

Recent research, \rca~\cite{chang2025firmrca} introduces 
automated bug diagnosis for re-hosting-based firmware fuzzing.  
\rca employs backward taint analysis to trace the propagation of faulty data and identify instructions  
related to the root cause of the crash.  
It also captures memory snapshots during crash replay to assist the analysis,  
enabling more precise fault localization in MCU firmware.

\begin{figure*}[t]
\setlength\abovedisplayskip{0pt}
\setlength\belowdisplayskip{0pt}
\centering
\includegraphics[width=\textwidth]{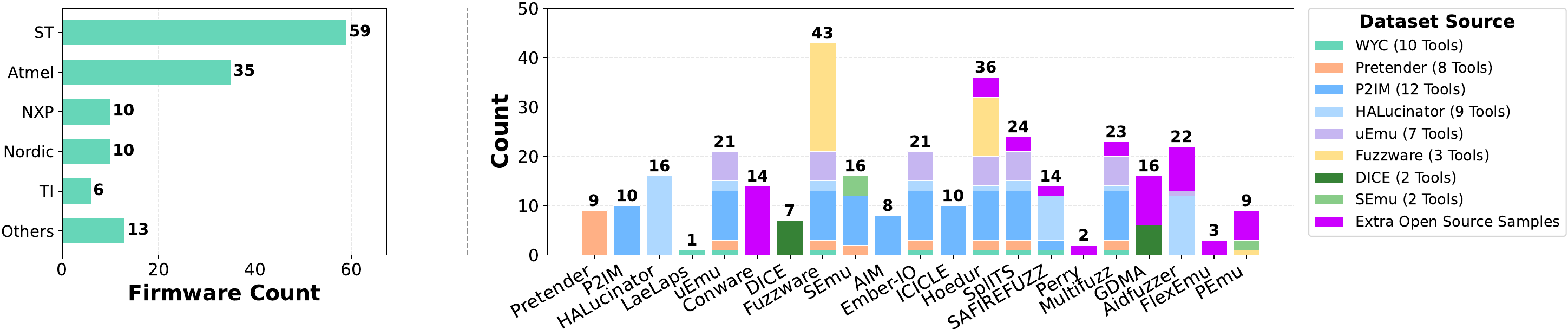}
\caption{The firmware samples used in evaluation of existing studies. The left
figure shows the distribution by vendor. The right figure illustrates the
composition of firmware sources used in each work.}
\label{fig:dataset}
\end{figure*}

\section{Large-Scale Empirical Study of SOTA Tools}

Current evaluations of emulation-based firmware analysis approaches remain
limited in scope and depth. Particularly, they rely on small, overlapping
datasets and assess only isolated components of the analysis pipeline. We
analyze the firmware datasets utilized in existing studies, excluding unit
tests. The results are shown in~\autoref{fig:dataset}. Compared to the diverse
datasets used in rehosting-based Linux firmware research~\cite{helmke2025mens},
datasets in monolithic MCU firmware research comprise only a few to several
dozen samples. These datasets often include demo programs from open-source
embedded OS communities like Zephyr and FreeRTOS, and firmware samples from
specific chip vendors such as STM32, Atmel, and NXP. Moreover, a few firmware
collections are popular among many tools. For example, the
\ppim~\cite{feng2020p2im} and WYC~\cite{muench2018you} datasets have been used by 12
and 10 tools, respectively. Among 325 unique samples, only 133 are used by only
one tool. On the one hand, this indicates a common benchmark in this field. On
the other hand, this means tools might be overfitting to the specific
characteristics of these samples, raising concerns about their applicability to
more diverse real-world firmware.

In this work, we present the first comprehensive, end-to-end evaluation of SOTA
emulation-based approaches using a large-scale, real-world MCU firmware dataset.
Our study systematically examines each stage of the automated firmware analysis
workflow, evaluating whether the output of one stage provides sufficient
information to enable the next. By feeding the results of each stage as inputs
to subsequent stages, we assess the true interoperability, robustness, and
completeness of existing emulation-based solutions.

\paragraph{Experiment Setup.} Building on top of the Ghidra API, our evaluation
pipeline comprises over 25,000 lines of Kotlin code. It automates batch
processing tasks, uses PostgreSQL to store test results and execution logs, and
supports modular parallel testing. All experiments were conducted on a server
with an Intel(R) Xeon(R) Platinum 8350C processor at 2.60 GHz, 512 GiB of
memory, and 15 TB of storage.

\subsection{Evaluation Scope} 
\label{sec:toolcov}

We present which tools are tested in our evaluation and which firmware samples
are used.

\paragraph{Tool Selection.}
Our goal was to run all the in-scope tools within a unified evaluation framework
using the same dataset. However, tools that were inaccessible or did not function
reliably in our environment were excluded. In stage 1, FirmXRay and FirmLine
from the literature are included, along with rbasefind~\cite{rbasefind}, a
popular brute-force base address scanner. These tools automatically recover
firmware base addresses and entry points from firmware binaries. Since no
other information can be obtained from existing tools, we excluded tools
requiring additional specifications, code-level information, ELF-formatted
memory maps, or other hardware-dependent information in stages 2 and 3, as
summarized in~\autoref{tab:depend}. 
The remaining candidates were \ppim~\cite{feng2020p2im},
DICE~\cite{mera2021dice} (\ppim plus DMA), AIM~\cite{feng2023aim},
\uemu~\cite{zhou2021automatic}, ICICLE~\cite{chesser2023icicle},
\fuzzware~\cite{tobias2022fuzzware}, \gdma~\cite{scharnowski2025gdma} (\fuzzware
plus DMA), \multifuzz~\cite{chesser2024multifuzz},
\aid~\cite{wang2025aidfuzzer}, and \hoedur~\cite{scharnowski2023hoedur}.
Additionally, \multifuzz's fuzzing front end incorporates ICICLE's enhancements,
while ICICLE's peripheral emulation depends on \fuzzware. We therefore excluded
ICICLE from the testing.
For Stage~4, \rca is the only tool that supports bug diagnosis within a firmware
emulation environment.

\paragraph{Firmware Collection.}
We used firmware datasets collected from \otacap (4,147 binary files, obtained
from the authors) and \firmline (21,507 binary files). Before conducting
experiments, we verified and filtered these binaries to (1) retain firmware
containing executable assembly code, and (2) identify their instruction set
architectures (ISAs). 

From the total of 25,654 collected files, we first removed 6,634 non-executable
files (\eg~\texttt{.txt}, \texttt{.mp3}). Among the remaining files, 1,421
binaries were in known executable formats, such as ELF. For unknown-format
binaries, we used Ghidra to disassemble each candidate across popular
ISAs~\cite{nonlinux2024}. Files for which Ghidra failed to identify at least ten
valid functions were considered non-firmware and removed from further analysis (see Appendix~\ref{app:threshold}).
We report the top ISAs identified among unknown-format binaries in
Appendix~\ref{app:isa}. In the end, \textbf{4,571} unique ARM Cortex-M firmware
binaries were identified, forming the input dataset for the analysis pipeline.
Using signature-based firmware recognition, we identify and categorize the
firmware based on their vendors. Details are shown in
Appendix~\ref{app:dataset-vendor}.

\subsection{Stage 1: Emulation Configuration Reconnaissance}
\label{sec:stage1results}

Since \firmline does not provide an entry point localization method, we adopted
the same assumption proposed in \firmxray to identify entry points. Because ground-truth base addresses and entry points are unavailable for black-box firmware, we performed a screening before feeding the firmware in Stage~2.
Using the Ghidra Emulator~\cite{ghidraEmulatorAPI}, we run each firmware with
the base address and entry point obtained from Stage~1. If the firmware executes
more than 10 unique instructions within the first 10,000 executed instructions
without terminating, we consider the base address and entry point valid. Passing
this screening does not guarantee correctness, but it removes configurations
that are obviously wrong, as explained in Appendix~\ref{app:threshold}.

\paragraph{Results.}
\autoref{tab:base-address-results} presents the base-address recovery results
for FirmXRay, FirmLine, and rbasefind across 4,571 firmware samples. For each
output, we further verified the results using the Ghidra Emulator. FirmXRay,
FirmLine, and rbasefind produced 2,468, 391 and 1,227 candidates that passed
initial screening, respectively. The reasons for no output and failed screening
are discussed in~\autoref{sec:gaps}.

\begin{table}[t]
\centering
\caption{Base-address recovery results in Stage 1.}
\label{tab:base-address-results}
\begin{adjustbox}{width=\columnwidth}
\begin{tabular}{|l|r|r|r|}
\hline
\multirow{2}{*}{\textbf{Tool}} &
\multirow{2}{*}{\textbf{No Output}} &
\multicolumn{2}{c|}{\textbf{Base Address Output}} \\
\cline{3-4}
& & \textbf{Pass Init. Screening} & \textbf{Fail Init. Screening} \\
\hline\hline
FirmXRay         & 19   & 2,468 & 2,084 \\\hline
FirmLine          & 2,953 & 391  & 1,227 \\\hline
rbasefind         & 913  & 1,227 & 2,431 \\
\hline
\end{tabular}
\end{adjustbox}
\end{table}

\subsection{Stage 2: Emulation Initialization}
\label{sec:stage2results}

Since \firmxray significantly outperforms \firmline, we used the
\textbf{2,468} samples whose \firmxray-generated base addresses and entry
points passed our verification for the subsequent Stage~2 experiments.

We evaluated nine tools from the tool-selection process on this dataset. To
maintain a consistent experimental methodology, we categorized these tools
according to the coupling between emulation (Stage~2) and security testing
(Stage~3).

\subsubsection{Decoupled Emulator–Fuzzer}

This category includes \ppim, DICE, AIM, and \uemu. \ppim and \uemu separate
peripheral modeling from fuzzing by introducing a dedicated preprocessing stage.
DICE and AIM depend on \ppim to generate MMIO response models before modeling
DMA or IRQ behavior.

Due to the extensive time required for the peripheral modeling phase, we imposed
a 1-hour termination limit for each tool. \ppim, AIM, and DICE failed on all 2,468
firmware images: 2,127 aborted immediately, and 341 remained in infinite
loops during peripheral modeling for over one hour. These failures arise from two
main issues. First, \ppim depends on a gnu-mcu-eclipse version of
QEMU~\cite{p2im_qemu_build}, which only supports specific ARM Cortex-M3/M4 MCU
series, such as STM32 and Kinetis, and is incompatible with other memory maps.
Second, the MMIO access classification patterns and modeling strategy employed
by \ppim are only effective for simple peripherals with straightforward usage,
as noted in prior studies~\cite{zhou2021automatic,tobias2022fuzzware}. Since AIM
and DICE are based on \ppim with enhancements for interrupts and DMA,
they suffer from similar limitations.

For \uemu, 651 images aborted without generating a peripheral model, and 1,522
failed to produce peripheral emulation models within 1 hour. 
\uemu successfully generated a peripheral modeling files for only 295 samples. This limitation occurs
because \uemu relies on concolic execution and requires at least one viable path
for peripheral modeling without reaching invalid states, infinite loops, or long
loops over 30,000 basic blocks. However, real-world firmware involves numerous
peripheral reads (each requiring a new symbol) and diverse checks on these
reads, leading to path explosion problems.

\subsubsection{Co-Designed Emulator–Fuzzer} 

This category comprises \fuzzware, \gdma, \multifuzz, \aid, and \hoedur. In
these systems, peripheral modeling is closely integrated with fuzzing:
peripheral interactions are inferred and modeled dynamically during the fuzzing loop.

To evaluate their emulation capability, 
we conducted fuzzing initialization tests using three 512-byte seeds: all zero bytes, all one bytes, and alternating 0/1 bytes, as utilized by GDMA, Hoedur, and MultiFuzz.
We considered emulation functional if all three seeds lead to successful
execution of the first 3 million basic blocks.
We classified any execution exceeding 3 million basic blocks but without any input consumption as a hang.

\begin{table}[t]
\centering
\caption{Failed emulation of 2,468 firmware images after fuzzing initialization. ``Total'' is the number of unique samples.
}
\label{tab:fuzz_init_failed_list}
\begin{adjustbox}{width=\columnwidth}
\begin{tabular}{|c|c|c|c|c|c|c|}
\hline
 & \textbf{\fuzzware} & \textbf{\gdma} & \textbf{\hoedur} & \textbf{\multifuzz} & \textbf{Total} & \textbf{Problem} \\
\hline\hline
UW      & 208  & 209  & \multirow{5}{*}{810} & 246                  & \multirow{5}{*}{831} & P1/2/3/5 \\
\cline{1-3}\cline{5-5}\cline{7-7}
UR      & 393  & 388  &                      & 483                  &  & P1/2/3/5 \\
\cline{1-3}\cline{5-5}\cline{7-7}
RV      & 1    & 1    &                      & 1                    & & P1/2/3/5 \\   
\cline{1-3}\cline{5-5}\cline{7-7}
UF      & 49   & 54   &                      & \multirow{2}{*}{46}  & & P1/2/3/5 \\
\cline{1-3}\cline{7-7}
EV      & 0    & 0    &                      &                      & & P1/2/3/5 \\
\hline
WV      & 15   & 16   & 17                   & 17                  & 27 & P1/2/3/5 \\
\hline
UI      & 194  & 199  & 0                    & 17                  & 215 & P4 \\ 
\hline
Hang    & 53   & 54   & 65                   & 62                  & 67 & P6 \\
\hline
Others   & 105   & 110   & 102                   & 123                 & 141 & P5/12 \\
\hline
\textbf{Total} & \textbf{997} & \textbf{1,006} & \textbf{983} & \textbf{969} & \textbf{1,123} & \\
\hline
\end{tabular}
\end{adjustbox}
\end{table}

As shown in~\autoref{tab:fuzz_init_failed_list}, 997, 1,006 and 969 samples
failed to emulate on \fuzzware, \gdma, and \multifuzz, respectively. 
\aid failed on 2,460 samples, with only 8 successful emulations.
Detailed failure statistics for each seed and tool are provided in
Appendix~\ref{app:init-fail}.
The failure
patterns are similar across tools. 
Manual analysis shows that incorrect base
addresses or entry points (P2/P3) commonly lead to unmapped accesses (UW/UR/UF),
inaccurate emulation (P4-P9) leads to other crashes, and long delay loops
produce hangs (P6), as discussed further in~\autoref{sec:gaps}.

\begin{table}[t]
\centering
\caption{Coverage distribution of firmware samples after one-hour fuzzing runs.}
\label{tab:coverage_dist}
\begin{adjustbox}{width=\columnwidth}
\begin{tabular}{|c|c|c|c|c|c|r|r|}
\hline
 & \textbf{\textless1} & \textbf{[1,5)} & \textbf{[5, 10)} & \textbf{[10, 20)} & \textbf{$\geq$20} & \textbf{Median} & \multicolumn{1}{c|}{\textbf{Mean}} \\
\hline\hline
\fuzzware         & 137 & 319 & 369 & 282 & 144 & 7.00\% & 9.19\% \\
\hline
\hoedur           & 127 & 325 & 274 & 329 & 249 & 7.92\% & 11.73\% \\   
\hline
\multifuzz        & 228 & 420 & 237 & 367 & 236 & 6.60\% & 10.23\% \\
\hline
\end{tabular}
\end{adjustbox}
\end{table}

\begin{table}[t]
    \centering
    \caption{Pairwise coverage comparison across fuzzers. A coverage difference of at least 1\% on the same sample is counted as a win; smaller differences are counted as ties.}
    \label{tab:pairwise_cov_comp}
    \begin{adjustbox}{width=\columnwidth}
    \begin{tabular}{|c|r|r|r|r|}
         \hline
         \multicolumn{1}{|c|}{\textbf{Pair (F vs L)}} & \multicolumn{1}{c|}{\textbf{F Win}} & \multicolumn{1}{c|}{\textbf{L Win}} & \multicolumn{1}{c|}{\textbf{Tie}} & \multicolumn{1}{c|}{\textbf{Total*}} \\
         \hline\hline
         \fuzzware vs \hoedur     & 46 (5.76\% avg) & 440 (8.32\% avg) & 589 & 1,075 \\
         \hline
         \fuzzware vs \multifuzz  & 138 (3.21\% avg) & 451 (8.05\% avg) & 583 & 1,172 \\
         \hline
         \hoedur vs \multifuzz     & 368 (3.72\% avg) & 108 (8.05\% avg) & 805 & 1,281 \\
         \hline
    \end{tabular}
    \end{adjustbox}
\flushleft
\scriptsize{
*: Total number of samples that both tools can support for fuzzing.
}
    
\end{table}

\subsection{Stage 3: Fuzzing and Runtime Results}
\label{sec:stage3results}

Due to the small number of samples supported by decoupled emulator–fuzzer tools
and \aid in Stage~2, we continued large-scale Stage~3 fuzzing with
\fuzzware, \gdma, \hoedur, and \multifuzz. We ran one-hour
fuzzing campaigns on all supported samples three times, recording
basic-block coverage, crashes, and hangs.
For fair comparison, because \multifuzz and \hoedur do not support DMA
enhancement, we compare \fuzzware with \multifuzz and \hoedur without the
\gdma plugin for coverage and crash results. Out of 35 cases involving recognized DMA models, only 27 exhibited slight differences between \gdma-enable \fuzzware and the original \fuzzware. Detailed \fuzzware fuzzing results are provided in
Appendix~\ref{app:gdma_comp}.

\paragraph{Coverage Results.} 
The coverage distribution for samples completing one-hour fuzzing is shown
in~\autoref{tab:coverage_dist}. The average fuzzing coverage for all evaluated tools ranges only from 9\% to 11\%.
Furthermore, only 426, 578, and 603 firmware samples achieved coverage exceeding
10\% for \fuzzware, \hoedur, and \multifuzz, respectively. We also directly
compared code coverage on the subsets of samples that each tool pair could
fuzz. As shown in~\autoref{tab:pairwise_cov_comp}, \hoedur and \multifuzz
generally achieved higher coverage than \fuzzware, indicating improved
fuzzing effectiveness. Among the three tools, \hoedur attained the highest
overall coverage, although the differences were relatively small in most
cases.

\paragraph{Crash Results.} 
Detailed crash results for the three fuzzers are shown
in Appendix~\ref{app:crash-hang}. Crashes occurred in more than one third of the
supported samples: 521/1,471 for \fuzzware, 681/1,499 for \multifuzz, and
619/1,485 for \hoedur. Manual analysis revealed that most crashes are false
positives, primarily caused by emulator limitations discussed
in~\autoref{sec:gaps} (P4--P9).
Although the underlying problems are similar across tools, the number of
reported crashes varies because the tools use different deduplication
strategies. \fuzzware lacks built-in deduplication, whereas \hoedur and
\multifuzz deduplicate specific crash types differently. As a result,
\fuzzware reports more crashes than \multifuzz and \hoedur. \hoedur
deduplicates UF/EV crashes based on the PC at the crash site, whereas
\multifuzz uses the last executed PC before the crash. Because the same
underlying issue can lead to different unmapped or permission-violating crash
PCs, especially for unmapped instruction fetches, \multifuzz reports fewer
crashes than \hoedur. We discuss the detailed differences between
\multifuzz's and \hoedur's deduplication strategies in
Appendix~\ref{app:dedup_strategy}.
We do not apply a single deduplication criterion across all tools because
only \fuzzware provides detailed crash information with complete traces,
whereas \hoedur and \multifuzz provide only deduplicated crash results.

\paragraph{Hang Results.}
\hoedur identified 27,987 hangs from 1,065 samples, whereas \multifuzz found
2,987 hangs from 1,301 samples. Because \hoedur does not deduplicate hangs,
its hang count is substantially higher than that of \multifuzz. Manual
analysis showed that current tools commonly hang on long delay loops, as
detailed in~\autoref{sec:gaps} (P6).

\subsection{Stage 4: Bug Diagnosis}
\label{sec:stage4results}
Given the large number of crashes from Stage~3, we randomly sample 100 crashes
that achieve at least 10\% code coverage for diagnosis. Crashes below this
threshold are more likely to result from incomplete initialization rather than
meaningful program execution; Appendix~\ref{app:threshold} details the
rationale.

\paragraph{Manual Validation.}
For each crash, we perform backward analysis from the crash site, following the
call stack and examining the relevant control- and data-flow dependencies. We
identify where the corrupted value is defined or propagated, and trace its
relationship to peripheral inputs and emulated hardware state. This allows us to
determine whether the crash originates from genuine firmware logic or from
emulation-induced behavior (P7/8/9). We compare the manually reconstructed
causal chain against that automatically identified by \rca.

\paragraph{Results.}
\rca analyzes the execution trace and reports a ranked list of candidate
root-cause instructions. Among the 100 crashes, \rca produces candidates for 98,
while the remaining two return no diagnostic result. Manual validation confirms
that all 98 diagnosed crashes are false crashes caused by P7--P9. For 79 cases,
\rca reports an instruction related to the immediate cause of the crash, but the
underlying cause lies earlier in the execution, where inaccurate emulation
corrupts the corresponding data or control flow. In the remaining cases, even
the reported candidate locations are unrelated to the reconstructed causal
chain. We further analyze these diagnostic limitations and their causes
in~\autoref{sec:gaps} (P10).

\section{Main Findings from the Empirical Study}
\label{sec:gaps}

\begin{figure*}[t]
\centering
\begin{minipage}[t]{0.63\textwidth}
  \vspace{0pt}
  \centering
  \includegraphics[width=\textwidth]{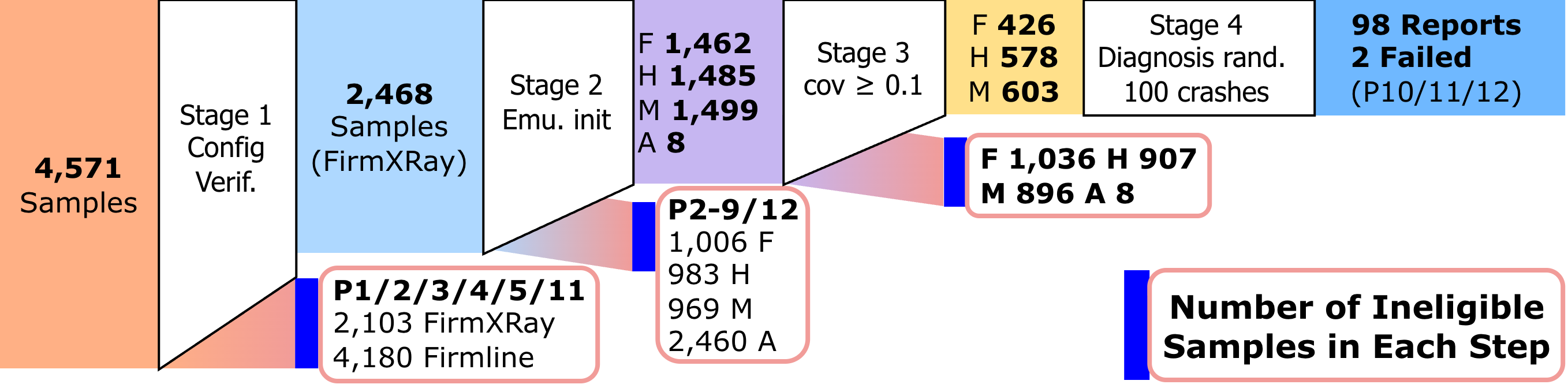}
  \caption{Sample attrition, stage-specific failures, and tool compatibility across the evaluation pipeline (F:Fuzzware, H:Hoedur; M:Multifuzz; A:AidFuzzer).}
  \label{fig:failworkflow}
\end{minipage}\hfill
\begin{minipage}[t]{0.34\textwidth}
  \vspace{0pt}
  \centering
  \begin{adjustbox}{width=\textwidth}
  \begin{tabular}{|l|l|l|}
    \hline
    \textbf{Problem} & \multicolumn{1}{c|}{\textbf{\fuzzware}} & \multicolumn{1}{c|}{\textbf{Multifuzz}} \\
    \hline\hline
    P4      & 12,315 (5.8\%)  & 1,549 (2.3\%) \\
    \hline
    P5      & 148,448 (70.1\%) & 38,192 (56.3\%) \\
    \hline
    P7      & 6,127 (2.9\%)   & 2,710 (4.0\%) \\
    \hline
    P8/9    & 12,201 (5.8\%)  & 18,561 (27.3\%) \\
    \hline
    P12     & -               & 41 (0.1\%) \\
    \hline
    Unconfirmed & 32,684 (15.4\%) & 6,840 (10.1\%) \\
    \hline
    Total   & 211,775 & 67,893 \\
    \hline
  \end{tabular}
  \end{adjustbox}
  \captionof{table}{The approximate prevalence of identified issues causing false crashes.}
  \label{tab:fp-problem-type}
\end{minipage}
\end{figure*}

We observe a steady drop in both effective firmware samples and applicable tools
along the evaluation pipeline, as visually summarized
in~\autoref{fig:failworkflow}. During Stage 2, some firmware fails before
fuzzing can start because the emulation environment cannot complete
initialization; we summarize these initialization failures and their associated
problems in~\autoref{tab:fuzz_init_failed_list}. For samples that enter Stage 3,
the evaluated tools report numerous crashes and hangs, most of which are false
positives under our manual validation and sampling-based analysis.
Table~\ref{tab:fp-problem-type} summarizes the number of false crashes associated
with each identified problem. We do not include \hoedur as its crash reports
provide limited diagnostic context. In Appendix~\ref{app:crash-hang}
and~\ref{app:problem-attribution}, we also provide more detailed breakdown of
crash types, crash counts, per-problem ratios within each type, and our heuristic
attribution methodology. 

Below, we discuss the main problems found from our empirical study. The hashes
of the numbered example firmware (\eg~\#2915) we refer to during discussion are
listed in Appendix~\ref{app:hashsample}. For readability, we rename variables
and functions in decompiled snippets while preserving the original instructions
and program logic.

\paragraph{P1: Insufficient Information Sources.}
Stage-1 base-address recovery tools generally infer the load address from
absolute pointers embedded in firmware, but differ in the sources they exploit.
FirmXRay considers IVT entries, global function pointers, and global string
pointers; FirmLine mainly relies on global function pointers; and rbasefind
relies on global string pointers. When a firmware contains no such artifacts,
these tools will fail. This limitation is particularly evident for FirmLine and
rbasefind, each of which relies on only one information source.

\paragraph{P2: Improper Candidate Weighting and Selection.}
When using absolute addresses to infer the base address, stage-1 tools solely
rely on the number of matches to weight candidates, leading to duplicated
occurrences to dominate the analysis. In firmware \#2915, 48 of the 57 IVT
entries point to the same address (\texttt{0xC069}). Consequently, \firmxray
infers \texttt{0x8A0C} as the base address because it best matches these
duplicates, even though it fails to resolve other candidates. Manual analysis
yields the correct base address \texttt{0x1000}.

Sometimes multiple candidates are produced, and existing tools, without more
evidence for further validation, tend to choose an arbitrary one. For example,
\firmxray produces two candidate base addresses (\texttt{0x1F000} and
\texttt{0x26000}) for firmware \#2797, and selects \texttt{0x1F000}. Manual
analysis confirms \texttt{0x26000} to be the correct one. Such improper
candidate weighting and selection account for many configuration reconnaissance
failures.

\paragraph{P3: Invalid IVT Placement Assumption.} 
\firmxray assumes that the IVT starts at the beginning of the firmware image and
uses its second entry (bytes 4-7), \ie~the reset handler, as the entry point.
This assumption holds for the majority of complete firmware images whose IVTs
are placed at the image start, such as certain Nordic and
Texas Instruments firmware, but this rule is not universal.

Sometimes, vendors may place some metadata or headers before the IVT, making
the fixed IVT-at-base assumption invalid. For example, sample \#1819, a
resistance probe from CorTalk~\cite{CorTalk}, starts with vendor-specific
metadata rather than an IVT (see~\autoref{lst:pher1819}). Interpreting bytes 4-7
as the reset-handler address therefore yields an invalid entry point, causing
immediate crashes and configuration-verification failure. Similarly, firmware
\#11175 (Marvell 88W8686) is an OTA image that begins with an OTA-download
header rather than the IVT.

\begin{lstlisting}[label={lst:pher1819},
abovecaptionskip=0pt, belowcaptionskip=0pt, escapechar=|,
caption={Header excerpt from sample \#1819}]
00000000   ds "R1ERj"
00000005   db 0x00,0x00,0x00,0xf4,0xcc
0000000a   db 0x01,0x00,0xbc,0xfd,0xe5
0000000f   ds "UApplication for RMU1ER"
\end{lstlisting}

\begin{tcolorbox}
[colback=cyan!10,
    colframe=white,
    width=\linewidth,
    boxsep=2pt,
    left=1pt,
    right=1pt,
    top=1pt,
    bottom=1pt]
\textbf{Takeaway 1}: 
A single method has to trade off between reliability and generality in recovering
base addresses and entry points.
Strict heuristics are often chip specific but yield high accuracy,
while general heuristics cover a wider range of firmware but may incur errors.
A mixture of general and chip-specific heuristics is desirable to cover a 
wider range of firmware with sufficient accuracy.
\end{tcolorbox}

\label{sec:S2P45}

\paragraph{P4: Unsupported ISA Instructions.}
\label{sec:P4}
As shown in~\autoref{fig:summary}, several firmware fuzzers use lightweight
CPU-emulation backends with incomplete ISA support. For example, \fuzzware is
based on Unicorn ver. 1.0.1, which does not support some instruction classes
available in QEMU, including floating-point and multimedia instructions. For example,
sample \#3741 is drone firmware that performs floating-point computation for
flight control. When execution reaches the floating-point instructions such as
\texttt{VLDR} and \texttt{VSTR}, the emulator crashes. As shown
in Table~\ref{tab:fp-problem-type}, this limitation causes 12,315 false crashes
across 74 samples in \fuzzware. In contrast, \hoedur, which uses QEMU as its
backend, does not exhibit this issue. A similar limitation appears in
\multifuzz, whose SLEIGH-based instruction translator lacks support for some
system instructions in Cortex-M, such as \texttt{SVC}. 

\paragraph{P5: Inaccurate Memory Layout.}
Memory layouts are very device specific and may contain multiple RAM, flash, or
external regions. Existing firmware fuzzers, without an accurate memory layout
of the target device, typically map a generic memory layout as shown
in~\autoref{tab:memorymap}. As a result, many legitimate accesses become
unmapped or conflicting accesses and are consequently reported as crashes. 

For example, in the sample \#227, the firmware accesses address
\texttt{0xA0001000} to interact with an external peripheral, and in the sample
\#12123, external RAM at \texttt{0x60000AFC} is legitimately accessed. These
addresses fall outside the generic memory layout and cause spurious memory
faults. Similarly, \aid maps only a part of the system-peripheral region
\texttt{0xE0000000-0xFFFFFFFF}, leading to an immediate crash for 99\% of the
tested samples when the firmware accesses other system peripherals.
We also observe that some firmware contains multiple fragments (\eg~bootloader,
application, radio controller, etc.), making the memory map more complex. For
example, in the sample \#3519, the image loads at \texttt{0x0} with a size of
131,072 bytes. However, its memory-fault ISR entry pointer is
\texttt{0x1001CA01}, indicating an additional executable region beyond the
loaded image.
Some firmware uses a high ROM base address that overlaps with the generic RAM
region at \texttt{0x20000000}. For instance, the \#2715 firmware uses a
ROM base address of \texttt{0x20001f40}, which conflicts with the emulated RAM
region. This ROM-RAM overlap results in memory access permission conflicts
within the emulator, causing initialization failures in \hoedur, \fuzzware, and
\multifuzz. We identified 25 such cases, as recorded under ``Others'' in
\autoref{tab:fuzz_init_failed_list}.

\begin{tcolorbox}
[colback=cyan!10,
    colframe=white,
    width=\linewidth,
    boxsep=2pt,
    left=1pt,
    right=1pt,
    top=1pt,
    bottom=1pt]
\textbf{Takeaway 2}: 
Robust CPU emulation necessitates extensive ISA support (P4).
Worse, architecture-level default configurations cannot capture the diversity of real-world memory layouts (P5).
In practice, accurately extracting memory maps from black-box, stripped firmware is challenging.
\end{tcolorbox}

\paragraph{P6: Absolute Speed Gap.}
Emulator execution is typically orders of magnitude slower than native execution
on real hardware. This speed gap is not a problem in general. However, it can
become a bottleneck when firmware contains long delay loops, which are common in
real-world firmware. For example, in firmware \#7658 (\autoref{lst:7658}) during
normal device initialization, \texttt{delay\_millis} invokes \texttt{count\_us}
999 times per \texttt{millis}; each call executes three basic blocks. When
\texttt{millis} equals 1,500, the delay alone executes over three million
basic blocks. Since \hoedur and \multifuzz treat executions beyond three million
basic blocks as a timeout, fuzzing terminates before reaching subsequent code.

\begin{lstlisting}[
    label={lst:7658},abovecaptionskip=0pt, belowcaptionskip=0pt, 
    escapechar=|, language=C, morekeywords={[1]{int, if, void, while}}, keywordstyle={[1]\color{magenta}},
caption={Delay-loop in the sample \#7658.}]
void delay_millis(int millis) {
    while (millis) {--millis; count_us(999);}
}
void  count_us (int count) {
    do --count; 
    while (count);
}
\end{lstlisting}

\begin{tcolorbox}
[colback=cyan!10,
    colframe=white,
    width=\linewidth,
    boxsep=2pt,
    left=1pt,
    right=1pt,
    top=1pt,
    bottom=1pt]
\textbf{Takeaway 3}: Long delay loops are common during peripheral
initialization. Emulating these delays faithfully hurts fuzzing efficiency.
\end{tcolorbox}

\paragraph{P7: Incorrect Interrupt Timing.}
On real devices, a peripheral interrupt is triggered only when three conditions
hold: (1) the interrupt is enabled in the interrupt controller, (2) the
corresponding peripheral interrupt-enable bits are set, and (3) the required
hardware event occurs. Most existing approaches (see~\autoref{tab:depend}),
including \fuzzware, \hoedur, and \multifuzz, only consider the first condition.
Therefore, all the enabled interrupts have a chance to be triggered. This can
invoke an ISR before the corresponding peripheral has been initialized, creating
infeasible executions and false crashes. For example, all crashes in sample
\#2938 occur upon the entry to ISR numbered 21, which dereferences an
uninitialized function pointer whose value resolves to \texttt{0x20003C00}.

\paragraph{P8: Inaccurate Peripheral Responses.} 
Existing peripheral-modeling approaches do not understand the semantic behavior
of the underlying hardware. Incorrect peripheral modelling can violate hardware
state and drive firmware into infeasible executions. For example, in the USART
ISR of the sample \#2691 (\autoref{lst:pher2691}), \texttt{frameLen} records the
amount of received data and should be updated only upon the corresponding
receiving event. However, existing tools unconditionally trigger the ISR by
satisfying checks on status registers. Consequently, \texttt{frameLen} may
remain zero while the branch is taken, causing \texttt{idx} to exceed
\texttt{rxBuf} and eventually triggering an invalid function-pointer
dereference. In the sample \#1179, incorrect emulated values cause the firmware
to remain in the loop until a timeout occurs. In the sample \#15998, an
unexpected clock-control register value causes an immediate persistent hang.

\begin{lstlisting}[
    label={lst:pher2691},
    abovecaptionskip=0pt, 
    belowcaptionskip=0pt, 
    escapechar=|, 
    language=C,
    morekeywords={[1]{int, if, void}},
    keywordstyle={[1]\color{magenta}},
caption={USART ISR in the sample \#2691.}]
int IRQ_handler(void) {
   if((MEM[0x40002304]&4)!=0 && MEM[0x40002108]){ 
     *(rxBuf + idx++) = MEM[0x40002518]; |\label{line:2691store}|
   if (frameLen == idx) { |\label{line:2691condition}|
     frameLen = 0; ...
\end{lstlisting}

\paragraph{P9: Incomplete Peripheral Modeling.} 
Existing models classify registers as configuration registers (CRs), status
registers (SRs), or data registers (DRs). Beyond producing inaccurate values
(P8), this abstraction cannot represent many register semantics found in real
devices. In complex peripherals such as USB, Ethernet, and radio controllers,
register values may encode addresses, offsets, descriptors, or other metadata
used to access peripheral memory. Simplifying peripheral register usage to broad
CR/SR/DR roles can miss richer hardware semantics and generate emulation-induced
false crashes.

For example, in the sample \#18803, GDMA models DMA incorrectly, treating a
pointer as DMA data. By supplying fuzzing input to the pointer, the firmware
eventually crashes when it fetches an invalid instruction. In the sample \#820
(\autoref{lst:pher820}), address \texttt{0x40005C50} corresponds to the USB
Buffer Table Address (\texttt{USB\_BTABLE}) register, whose bits 15--3 specify
the buffer-table location. However, \fuzzware treats it as a generic input and
supplies unconstrained fuzzing data. In one crashing execution, the register
returns \texttt{0xF3F3F3F3} at~\autoref{line:820read}, which is then interpreted
as address-related metadata and leads to an out-of-bounds read at
\texttt{0x27E84822} at~\autoref{line:820invalid}. 

\begin{lstlisting}[
    label={lst:pher820},
    abovecaptionskip=0pt, 
    belowcaptionskip=0pt, 
    escapechar=|,
    alsoletter={\#},
    morekeywords={[1]{ldr, addw, lsls, ldr}},
    keywordstyle={[1]\color{instr}},
    morekeywords={[2]{r2, r3}},
    keywordstyle={[2]\color{reg}},
    morekeywords={[3]{\#0x50, \#0x43c, \#0x1}},
    keywordstyle={[3]\color{pseudo}},
    caption={Peripheral-derived memory addresses in the sample \#820.}
]
ldr        r3,[DAT_08028d9c] ;= 40005C00h
ldr        r2,[r3,#0x50]     ;=>DAT_40005c50 |\label{line:820read}|
addw       r3,r3,#0x43c
lsls       r2,r2,#0x1
ldr        r3,[r3,r2]   |\label{line:820invalid}|
\end{lstlisting}

\begin{tcolorbox}
[colback=cyan!10,
    colframe=white,
    width=\linewidth,
    boxsep=2pt,
    left=1pt,
    right=1pt,
    top=1pt,
    bottom=1pt]
\textbf{Takeaway 4}: Low-fidelity heuristic peripheral I/O modeling
(P7/P8/P9) frequently produces false crashes. Improving peripheral emulation
fidelity remains difficult, especially in black-box settings.
\end{tcolorbox}

\paragraph{P10: Shallow Root-Cause Diagnosis.}
Although \rca can backtrack crash-related data dependencies and report candidate
root-cause locations, it cannot provide sufficient context to validate or rank
their causal relevance. For example, the sample \#3349 crashes on a NULL-pointer
dereference at \texttt{0xC}. \rca traces the invalid pointer back to a firmware
memory-initialization site. Manual analysis, however, shows that the actual root
cause occurs earlier: an inaccurate peripheral response drives execution down an
incorrect branch, skipping the function that should initialize the pointer and
eventually causing the invalid dereference. Technically, \rca lacks fine-grained
control-flow and input-context analysis to explain how an incorrect value or
initialization state was reached.

\begin{tcolorbox}
[colback=cyan!10,
    colframe=white,
    width=\linewidth,
    boxsep=2pt,
    left=1pt,
    right=1pt,
    top=1pt,
    bottom=1pt]
\textbf{Takeaway 5}: 
Accurate diagnosis of firmware crashes remains challenging (P10), as the
root cause may involve intertwined data- and control-flow dependencies,
compounded by the emulation error itself.
\end{tcolorbox}

\label{sec:engP10P11}

\paragraph{P11: Poor Scalability.}
Our evaluation reveals substantial time and memory overheads across the stages.
In Stage~1, \firmxray and \firmline perform near-brute-force base-address
inference. In several cases, \firmline spawned excessive subprocesses and
consumed over 200\,GB of memory (see Appendix~\ref{app:firmxray_bad_perf}). In
Stage~2, most decoupled Emulator–Fuzzer tools failed to complete the
peripheral modeling phase within one hour for most samples. In Stage~4, \rca
performs taint-based root-cause analysis and suffers from taint explosion on
long crash traces; as trace length increases, both analysis time and memory
consumption grow almost exponentially.

\paragraph{P12: Implementation Limitations.}
Beyond design issues such as peripheral modeling limitations, we identify additional
failures stemming from incomplete or incorrect tool implementations. These cases
are categorized as ``Others'' in Stage~2 (\autoref{tab:fuzz_init_failed_list})
and Stage~3 (Appendix~\ref{app:crash-hang}). For example, \hoedur, \fuzzware,
and \multifuzz do not implement ARM hardware reset through the
\texttt{VECTRESET} field of \texttt{AIRCR}, causing execution to abort when
firmware requests a reset. This limitation affects 94 samples. \multifuzz
occasionally misdecodes ARM/Thumb code and does not correctly handle
instruction-set mode transitions. When execution switches to Thumb code,
incorrect decoding can corrupt processor state. A detailed breakdown of these
failures is provided in Appendices~\ref{app:other-crash} and~\ref{app:init-fail}.

\paragraph{P13: Miscellaneous.}
First, most existing emulators predominantly support ARM Cortex-M, with limited
support for non-ARM architectures. Second, many tools require prerequisite
information often unavailable for black-box firmware. As shown in the
``Dependence Information'' column, only 12 out of 21 Stage-2 tools can function
without additional information beyond the base address and entry point. Third,
Stage-3 tools are frequently tied to specific Stage-2 tool implementations,
restricting interoperability between emulators and downstream analysis tools.
These architectural, informational, and integration constraints limit the
applicability of current techniques.

\begin{tcolorbox}
[colback=cyan!10,
    colframe=white,
    width=\linewidth,
    boxsep=2pt,
    left=1pt,
    right=1pt,
    top=1pt,
    bottom=1pt]
\textbf{Takeaway 6}: The practical utility of firmware emulation tools depends
not only on their core techniques, but also on scalable and robust
implementations, minimal prerequisite information, broad architecture support,
and interoperability across workflow stages.
\end{tcolorbox}

\section{Future Directions}

Based on the pitfalls in~\autoref{sec:gaps}, we outline six directions (D1--D6)
for reliable, scalable, and comparable analysis of real-world MCU firmware.

\paragraph{D1. Emulation Configuration Reconnaissance.}
Stage~1 remains underexplored but is essential for re-hosting stripped
black-box firmware. Rather than relying on individual heuristics, future methods
should jointly infer the ISA, execution mode, memory layout, entry point, HAL
functions, MCU model, and other platform metadata from complementary structural
and semantic evidence. This broader recovery is necessary because existing
approaches require different prerequisites and coarse configurations can cause
false failures. Explicitly representing uncertainty would further allow
downstream tools to refine or reject ambiguous configurations before analysis.

\paragraph{D2. Extensible Emulation Frameworks.}
Emulation advances remain difficult to reuse because tools bind peripheral
models, DMA and interrupt support, and fuzzing logic to incompatible back ends;
for example, \gdma is currently applicable only to \fuzzware, whereas AIM and
DICE target \ppim. A common substrate should separate CPU execution, hardware
modeling, event scheduling, and analysis through stable APIs for memory maps,
MMIO, interrupts, DMA, state inspection, tracing, and feedback.
LibAFL-QEMU~\cite{malmain2024libafl} and ICICLE~\cite{chesser2023icicle} are
promising foundations, but need broader architecture support, complete
interfaces, and competitive performance.

\paragraph{D3. High-Fidelity Automatic Peripheral Modeling.}
Heuristic models may preserve execution while misrepresenting register
semantics, peripheral responses, and interrupt timing (P7--P9). Higher-fidelity
approaches such as FlexEmu and \semu require HAL code or MCU specifications that may be unavailable for black-box firmware. Future work should construct reusable
peripheral-behavior repositories from specifications, drivers, and hardware
traces, then match or synthesize context-sensitive models using static and
runtime evidence. Models should expose both their assumptions and uncertainty,
not merely whether they keep firmware running.

\paragraph{D4. Emulation-Based Postmortem Analysis.}
Inaccurate emulation creates false crashes, while genuine MCU bugs may have
delayed or weak symptoms~\cite{muench2018you}. Existing tools provide limited
replay, deduplication, and causal diagnosis, and P10 shows that data flow alone
is insufficient. Postmortem systems should combine inputs, control and data
dependencies, peripheral and interrupt events, and emulator state in replayable
traces to distinguish firmware bugs from modeling artifacts, group related
failures, and identify both the triggering operation and earlier state
corruption.

\paragraph{D5. Realistic and Unified Benchmarks.}
MCU benchmarks under-represent peripheral diversity and favor simple MMIO
polling over interrupt-, DMA-, state-, and timing-dependent behavior. Unified
benchmarks should span architectures, vendors, models, RTOSes, and peripheral classes,
with ground truth for configuration, hardware-visible behavior, vulnerabilities,
triggering inputs, and root causes. Fidelity metrics should measure trace and
state agreement, not only boot or execution progress. The domain still lacks a
widely adopted counterpart to LAVA-M~\cite{dolan2016lava};
FirmRebugger~\cite{duong2026firmrebugger} is a promising step toward
fuzzing ground-truth evaluation.

\paragraph{D6. Standardized Evaluation Criteria.}
MCU firmware analysis lacks consistent criteria across the pipeline.
Comparability requires common reporting of prerequisites, assumptions, manual
effort, resource use, and tool versions. Emulation success should be evaluated
with firmware-dependent fidelity levels, rather than simply whether firmware
runs without exceptions,
following the
spirit of Greenhouse~\cite{tay2023greenhouse}. Fuzzing  should also adopt
consistent definitions and reporting for initialization, throughput, coverage,
timeouts, crashes, hangs, deduplication, and false-positive validation, consistent with
general-purpose fuzzing practices~\cite{schloegel2024sok}.
Machine-readable
configurations and outputs would further improve reproducibility.

\begin{table*}[t]
\centering
\caption{Firmware Emulation Revisited: Progress on Past Challenges, Persistent Issues (P), and Emerging Challenges (E).}
\label{tab:prior-sok-relation}
\scriptsize
\renewcommand{\arraystretch}{1.02}
\setlength{\tabcolsep}{2pt}

\begin{tabularx}{\textwidth}{
|>{\centering\arraybackslash}p{0.045\textwidth}
|>{\raggedright\arraybackslash}p{0.40\textwidth}
|>{\centering\arraybackslash}p{0.09\textwidth}
|>{\raggedright\arraybackslash}X
|>{\centering\arraybackslash}p{0.06\textwidth}
|>{\centering\arraybackslash}p{0.06\textwidth}|
}
\hline
\textbf{Stage} &
\textbf{Conclusion} &
\textbf{Reference} &
\textbf{Takeaway/Evidence} &
\textbf{Problem} &
\textbf{Direction} \\
\hline\hline

S1 &
\textbf{Challenge:} Emulation requires firmware metadata
(memory layout, ISA, entry point (EP), etc.). &
P (\cite{wright2021challenges}:\S7;
\cite{fasano2021sok}:\S3.1--3.2) &
Takeaway 1 &
P1--P3 &
D1 \\

&
$\hookrightarrow$ \textit{Method:} Infer base addresses from function
addresses via statistical voting. &
\cite{wright2021challenges}:\S7.3 &
$\hookrightarrow$\textit{Evidence:} Sparse candidates and equal-weight voting can cause
incorrect inference. &
P1--P2 &
 \\

&
$\hookrightarrow$ \textit{Method:} Infer EP from hardware conventions
(e.g., Cortex-M initial PC at vector-table offset 0x4). &
\cite{wright2021challenges}:\S7.4 &
$\hookrightarrow$\textit{Evidence:} Unknown hardware or customized image layouts invalidate
fixed-offset assumptions. &
P3 &
 \\
\hline

S2 &
\textbf{Challenge:} Peripheral diversity exceeds the number of supported
peripherals in any emulation system. &
P (\cite{wright2021challenges}:\S8.1.1;
\cite{fasano2021sok}:\S4.4,4.6) &
Takeaway 4 &
P7--P9,P13 &
D3 \\

S3 &
$\hookrightarrow$ \textit{Method:} Use fuzzer-based peripheral modeling
to enable vulnerability discovery. &
\cite{wright2021challenges}:\S8.1.1 &
$\hookrightarrow$\textit{Evidence:} Low-fidelity modeling can produce many false crashes. &
P8,P9 &
 \\

S2 &
$\hookrightarrow$ \textit{Method:} Trigger IRQs manually, invoke handlers,
or patch automatic triggers. &
\cite{wright2021challenges}:\S8.2.6 &
$\hookrightarrow$\textit{Evidence:} Incorrect IRQ timing causes failures; existing triggering
strategies lack generality. &
P7 &
 \\

S2 &
$\hookrightarrow$ \textit{Method:} Handle peripheral interactions and DMA through
Hardware-in-the-loop or HAL abstractions. &
\cite{wright2021challenges}:\S8.1.1,8.2.5;
\cite{fasano2021sok}:\S5.2--5.3 &
$\hookrightarrow$\textit{Evidence:} Hardware/HAL-dependent approaches do not readily
generalize to black-box firmware. &
P13 &
 \\
\hline

S2 &
\textbf{Challenge:} Locate data, code, and peripheral regions. &
P (\cite{wright2021challenges}:\S7.5) &
Takeaway 2 &
P5 &
D1 \\

&
$\hookrightarrow$ \textit{Method:} Use built-in memory handlers or custom
modules for memory interactions. &
\cite{wright2021challenges}:\S8.1.2,8.1.4 &
$\hookrightarrow$\textit{Evidence:} Requires layout/behavior knowledge often unavailable
in black-box firmware. &
P5 &
 \\
\hline

S2 &
\textbf{Challenge:} Full ISA support for CPU emulation. &
E &
Takeaway 2 &
P4 &
D2 \\
\hline

S3 &
\textbf{Challenge:} Long delay loops hinder fuzzing. &
E &
Takeaway 3 &
P6 &
D2 \\
\hline

S2/3 &
\textbf{Challenge:} Verifying re-hosting fidelity remains difficult. &
P (\cite{wright2021challenges}:\S9.2;
\cite{fasano2021sok}:\S3.4,7.4) &
Takeaway 6 &
- &
D5--D6 \\
\hline

S4 &
\textbf{Challenge:} Accurate bug diagnosis under emulated execution. &
E &
Takeaway 5 &
P10 &
D4 \\
\hline

S1/2/3/4 &
\textbf{Challenge:} Performance, implementation, and applicability limitations. &
E &
Takeaway 6 &
P11--P13 &
D5--D6 \\
\hline
\end{tabularx}
\flushleft
\scriptsize{Some conclusions are simplified while retaining their original meaning. Section numbers following references indicate the corresponding sections in~\cite{wright2021challenges,fasano2021sok}.}

\end{table*}

\section{Discussion and Limitations}

\subsection{Progress on Past Challenges, Persistent Issues, and Emerging Challenges} 

After five years of research, we revisit the challenges and mitigation
techniques identified by prior SoKs~\cite{wright2021challenges,fasano2021sok} in
light of our large-scale evaluation. We summarize our findings in
\autoref{tab:prior-sok-relation}, classifying each challenge as either
\textbf{P} (Persistent) if it remains valid or \textbf{E} (Emerging) if it is
newly identified in this study. Unfortunately, despite significant advancements
in mitigation techniques, none of the challenges identified by prior SoKs is
fully resolved. 

We revisit the mitigation methods proposed in prior SoKs. Our experiments
provide concrete evidence as to why they cannot fully address these challenges,
especially for black-box firmware. Under each challenge, our table lists the
corresponding mitigation methods and the evidence from our evaluation that
highlights their limitations. Additionally, we identify challenges overlooked in
prior SoKs. Notably, post-fuzzing analysis for MCU firmware was largely
non-existent before. Our evaluation uncovers difficulties in distinguishing
genuine firmware defects from emulation-induced crashes and in reliably
diagnosing their root causes. Our pipeline perspective also reveals cross-stage
interface and dependency issues that are difficult to detect when emulation,
fuzzing, and diagnosis are studied independently. The corresponding takeaways,
problems, and future directions are linked through identifiers in the table.

\subsection{Limitations of This Study}
\label{sec:limited-finding-scope}

Our study is framed within a pipeline-oriented approach for real-world black-box
firmware. Many tools that are not compatible with this pipeline, \eg~those that
require specific inputs or configurations that are unavailable, are excluded from
our evaluation. For example, some high-fidelity peripheral modeling tools like
\semu~\cite{zhou2022your}, HALucinator~\cite{clements2020halucinator}, and
Perry~\cite{chong2024afriend} require hardware manuals or source-level
HAL/driver code. For black-box firmware, this is very challenging. P4-P9 might
be biased specifically because of this limitation.

Our pipeline does not test how these tools deal with non-ARM firmware. Limiting
the scope to ARM enables a more focused evaluation of the emulation technique
itself. However, this also means our findings are not directly applicable to
other architectures.
 
Lastly, our evaluation involves many empirical observations. For example, we use
a threshold of 10 valid functions to distinguish firmware from non-code blobs.
While we have striven to ensure objectivity, some of our conclusions may be
influenced by subjective judgments. We have attempted to mitigate this by
providing detailed evidence and rationales for our assessments in
Appendix~\ref{app:threshold}, but we acknowledge that different researchers
may interpret the same data differently.

\section{Conclusion}

In this SoK, we systematized MCU firmware emulation-based research within a
unified four-stage dynamic analysis workflow covering emulation
configuration reconnaissance, emulation, fuzzing, and bug diagnosis. We then
conducted the first large-scale empirical evaluation of SOTA approaches on
real-world MCU firmware. Through statistical analysis and detailed case
studies, we exposed recurring pitfalls that limit current tools, including
fragile configuration inference, incomplete emulation support, low-fidelity
peripheral modeling, unreliable crash diagnosis, and engineering limitations
that hinder large-scale use. We hope these findings and future directions
provide a foundation for more robust, comparable, and practical firmware
dynamic analysis systems.

\section*{Ethical Considerations}

Our study applies existing emulation-based analysis and fuzzing tools to
real-world firmware and reports tens of thousands of crashes. As shown in our
analysis, many crashes result from emulation inaccuracies and are false
positives. However, manually validating every crash is infeasible, leaving some
unverified crashes that could be false positives. To date, we have not confirmed
any previously unknown vulnerability. If a real vulnerability is identified, we
will promptly notify the affected vendor and follow responsible disclosure
practices. To prevent misuse, we only release aggregate statistics for
unverified crashes. Some firmware may be subject to copyright or redistribution
restrictions. We do not redistribute firmware images directly as is done in the
current practice; researchers can reconstruct the benchmark using authorized
sources and the provided hash values for the firmware.

\section*{Open Science and Artifact Availability}

To enable future research, we release the following artifacts at
\url{https://github.com/IoTS-P/LargeRehostingTesting}.

\paragraph{Evaluation Framework.}
The source code of our automated pipeline-based firmware evaluation framework.

\paragraph{Evaluated Tools and Configurations.}
Snapshots of all evaluated tools, including tool versions, build instructions,
patches, and per-tool configurations, subject to their respective licenses.

\paragraph{Evaluation Results.}
Final results and intermediate outputs in each pipeline stage:
\begin{itemize}
    \item \textbf{Stage 1:} Recovered base addresses and entry points,
    configuration-verification results, and problem categories for failed
    verification. 
    
    \item \textbf{Stage 2:} For each tool-firmware combination, the
    initialization results, the used seeds, and problem categories for
    unsuccessful cases.
    
    \item \textbf{Stage 3:} For each tool-firmware combination, the fuzzing
    results, including coverage, deduplicated crash counts, and problem
    categories for unsuccessful cases.
    
    \item \textbf{Stage 4:} Diagnostic outputs from \rca and the corresponding
    manual verification results.

\end{itemize}

\paragraph{Details of the Empirical Study.}
The detailed evaluation results and additional cases for each problem to support
the findings and conclusions discussed in~\autoref{sec:gaps}.

\paragraph{Firmware Dataset Reconstruction Methods.}
FirmLine~\cite{balgavy2024firmline} firmware is publicly available, whereas
OTACap~\cite{nino2024unveiling} dataset is authorized by the original authors
under a non-disclosure agreement. To uniquely identify and reconstruct the
evaluated benchmark, we provide MD5 hashes, source-dataset identifiers, and
mappings. Researchers can use these to retrieve the original firmware from the
respective sources. We also provide firmware metadata recoverable from prior
techniques and collected artifact signatures, including vendor, OS, and
device/MCU model information where identifiable.

\section*{Acknowledgment}
We sincerely appreciate our shepherd and all the anonymous reviewers for their insightful and valuable feedback. This work was supported by National Natural Science Foundation of China (NSFC) grant (62202188).

\bibliographystyle{IEEEtran}
\bibliography{refs}

\appendix
\raggedbottom

\subsection{Additional Literature Scope}
\label{app:litscope}

\begin{itemize}
    \item European Symposium on Research in Computer Security (ESORICS)
    \item Annual Computer Security Applications Conference (ACSAC)
    \item ACM Asia Conference on Computer and Communications Security (Asia CCS)
    \item International Symposium on Recent Advances in Intrusion Detection (RAID)
    \item IEEE/ACM International Conference on Automated Software Engineering (ASE)
    \item ACM Joint European Software Engineering Conference and Symposium on the Foundations of Software Engineering (ESEC/FSE)
    \item ACM SIGSOFT International Symposium on Software Testing and Analysis (ISSTA)
    \item International Conference on Software Engineering (ICSE)
    \item International Conference on Dependable Systems and Networks (DSN)
    \item European Symposium on Security and Privacy (EuroS\&P)
    \item IEEE Transactions on Dependable and Secure Computing (TDSC)
    \item IEEE Transactions on Information Forensics and Security (TIFS)
\end{itemize}

\subsection{Dataset ISA Identification}

\label{app:isa}

\begin{table}[H]
  \centering
  \caption{Summary of ISA identification for the collected dataset.}
  \label{tab:isacount}
  \begin{adjustbox}{width=0.8\columnwidth}
  \begin{tabular}{|l|l|r|}
    \hline
    \textbf{Arch} & \textbf{ISA} & \textbf{Count}\\
    \hline\hline
    \multirow{4}{*}{ARM}    & ARM/Thumb, little endian, 16/32 bit & \textbf{4,571}\\
                            & ARM/Thumb, Big endian, 32 bit & 1\\
                            & AARCH64, Little endian, 64 bit & 210\\
                            & AARCH64, Big endian, 64 bit & 8\\
    \cline{1-3}
    \multirow{2}{*}{MIPS}   & MIPS, Little endian, 32/64 bit & 10\\
                            & MIPS, Big endian, 32/64 bit & 13\\
    \cline{1-3}
    RISC-V                  & RISC-V, Little endian, 32/64 bit & 177\\
    \cline{1-3}
    \multirow{2}{*}{ESP32}  & XTensa, Little endian, 32 bit & 751\\
                            & XTensa, Big endian, 32 bit & 43\\
    \cline{1-3}
    x86                     & x86, Little endian, 32/64 bit & 82\\
    \cline{1-3}
    Total                   & &5,866\\
    \hline
  \end{tabular}
  \end{adjustbox}
\end{table}

\subsection{MCUs and MCU Vendors of Firmware Samples Used in Previous Works}
\label{app:previous_work_vendors}

\begin{table}[H]
\centering
\caption{MCU and Vendor Statistics in Previous Works (fuzzing-only; 133 firmware in total)}
\label{tab:firmware_mcu_stat}
\begin{adjustbox}{width=\columnwidth}
\begin{tabular}{|c|l|r||c|l|r|}
\hline
\textbf{Vendor} & \textbf{MCU Model} & \textbf{Count} & \textbf{Vendor} & \textbf{MCU Model} & \textbf{Count} \\
\hline
\hline
\multirow{19}{*}{STM32} & STM32F103 & 12 & \multirow{6}{*}{Atmel} & SAM3X & 18 \\
 & STM32F429 & 11 &  & SAMR21 & 9 \\
 & STM32F479 & 7 &  & SAM4E & 3 \\
 & STM32L152 & 7 &  & SAM4S & 3 \\
 & STM32F072 & 3 &  & SAM4L & 1 \\
 & STM32F303 & 3 &  & SAMD21 & 1 \\
\cline{4-6}
 & STM32F3 & 2 & \multirow{4}{*}{NXP} & K64 & 7 \\
 & STM32F767 & 2 &  & LPC1549 & 1 \\
 & STM32L475 & 2 &  & LPC1768 & 1 \\
 & STM32F207 & 1 &  & LPC1837 & 1 \\
\cline{4-6}
 & STM32F405 & 1 & \multirow{2}{*}{Nordic} & nRF52840 & 9 \\
 & STM32F407 & 1 &  & nRF52832 & 1 \\
\cline{4-6}
 & STM32F411 & 1 & TI & CC2538 & 6 \\
\cline{4-6}
 & STM32F466 & 1 & Maxim & MAX32600 & 3 \\
\cline{4-6}
 & STM32F4xx & 1 & Renesas & RA4W1 & 2 \\
\cline{4-6}
 & STM32F769 & 1 & Silicon & EFM32 & 3 \\
\cline{4-6}
 & STM32L431 & 1 & Infineon & CY8C63 & 1 \\
\cline{4-6}
 & STM32L432 & 1 & Holtek & HT32F52342 & 1 \\
\cline{4-6}
 & STM32WL55 & 1 & Unknown &  & 3 \\
\hline
\end{tabular}
\end{adjustbox}
\end{table}

\subsection{Vendor and Source Distribution of Our Dataset}
\label{app:dataset-vendor}
We identify firmware vendors using two complementary signature sources: artifact signatures provided by OTACap~\cite{nino2024unveiling} and our keyword-based signatures. Table~\ref{tab:vendors} reports both the number of samples identified by each signature source and the distribution of samples across the OTACap and FirmLine source datasets. In total, 3,869 samples originate from OTACap and 702 from FirmLine, totaling 4,571 samples. ``Unidentified'' denotes samples for which vendor information cannot be recovered using the available signatures.

\begin{table}[H]
\centering
\caption{Firmware identification by signature and source.}
\label{tab:vendors}
\label{tab:vendor-source}
\begin{adjustbox}{width=\columnwidth}
\begin{tabular}{|l|r|r|r|r|r|}
\hline
\multirow{2}{*}{\textbf{}} & \multicolumn{2}{c|}{\textbf{by Signature}} & \multicolumn{2}{c|}{\textbf{by Source}} & \multirow{2}{*}{\textbf{Total}} \\
\cline{2-5}
 & \textbf{OTACap Sig.} & \textbf{Keywords} & \textbf{OTACap} & \textbf{FirmLine} & \\
\hline\hline
\textbf{Nordic} & 0 & 878 & 689 & 189 & 878 \\
\hline
\textbf{TI} & 165 & 15 & 173 & 7 & 180 \\
\hline
\textbf{STM32} & 0 & 53 & 41 & 12 & 53 \\
\hline
\textbf{ESP} & 0 & 37 & 37 & 0 & 37 \\
\hline
\textbf{Dialog} & 18 & 15 & 33 & 0 & 33 \\
\hline
\textbf{Realtek} & 0 & 15 & 14 & 1 & 15 \\
\hline
\textbf{Telink} & 0 & 15 & 6 & 9 & 15 \\
\hline
\textbf{Cypress} & 0 & 14 & 14 & 0 & 14 \\
\hline
\textbf{NXP} & 0 & 5 & 5 & 0 & 5 \\
\hline
\textbf{Silabs} & 0 & 4 & 2 & 2 & 4 \\
\hline
\textbf{Microchip} & 0 & 1 & 0 & 1 & 1 \\
\hline
\textbf{Unidentified} & - & - & 2,855 & 481 & 3,336 \\
\hline
\end{tabular}
\end{adjustbox}
\end{table}

\subsection{Hashes of Mentioned Firmware Samples}

The hashes of the real-world samples referenced in the paper are listed
in~\autoref{tab:sample}. 

\label{app:hashsample}

\begin{table}[H]
    \centering
    \caption{Hashes of the firmware samples discussed in this paper.}
    \label{tab:sample}
    \begin{adjustbox}{width=\columnwidth}
    \begin{tabular}{|r|l|l|}
         \hline
         \multicolumn{1}{|c|}{\textbf{No.}} & \multicolumn{1}{c|}{\textbf{MD5 Hash}} & \multicolumn{1}{c|}{\textbf{Source}} \\
         \hline\hline
         \#227 & 0a3d1bcfbf1975c6f9251d94b5d6068c & OTACAP \\
         \hline
         \#820 & ccb629ef7798b16ce643c93c98def4da & OTACAP \\
         \hline
         \#1179 & cf7e5ffce5627d94782f775cebe37d5e & OTACAP \\
         \hline
         \#1819 & ffd04cf806f9eb938b91d2125dd0cf4f & OTACAP \\
         \hline
         \#2691 & 6a2c995e4ff7de99b1666a57141edb59 & OTACAP \\
         \hline
         \#2715 & 39f97e664aa557a79742d3cc396cf356 & OTACAP \\
         \hline
         \#2797 & 9c69319bd19d6e70c52dbf52706c306c & OTACAP \\
         \hline
         \#2915 & 7d798d66e6f6b7a83339cdc7229e25d3 & OTACAP \\
         \hline
         \#2938 & d23dcc6078f806bcf861c3774882a810 & OTACAP \\
         \hline
         \#3349 & 8ac0e15548f08aa0bade6d1e39def173 & OTACAP \\
         \hline
         \#3519 & 761c7650abcb516c1222115c3e60a673 & OTACAP \\
         \hline
         \#3741 & 2aff89ee5b54eb5e0dc01a3afa2489f8 & OTACAP \\
         \hline
         \#7658 & 98ebfc674e848b9fddf0c8e4546684b4 & Firmline \\
         \hline
         \#11175 & b410ae085ca5b9e539fa82b785becdff & Firmline \\
         \hline
         \#12123 & afb512988a3bb8c5f5c1b0d56f444ec7 & Firmline \\
         \hline
         \#15998 & 2b9c07679b3afb37f44fd0aeec5af533 & Firmline \\
         \hline
         \#18803 & b00141539d039261267c6c7c0fcade7f & Firmline \\
         \hline
    \end{tabular}
    \end{adjustbox}
\end{table}

\subsection{Resource-Intensive \firmline Failure Cases}
\label{app:firmxray_bad_perf}

\begin{table}[H]
    \centering
    \caption{Representative \firmline failure cases with high time and memory costs.}
    \label{tab:costy_firmxray}
    \begin{adjustbox}{width=0.9\columnwidth}
    \begin{tabular}{|r|r|c|r|}
         \hline
         \multicolumn{1}{|c|}{\textbf{No.}} & \multicolumn{1}{c|}{\textbf{File Size}} & \multicolumn{1}{c|}{\textbf{Runtime}} & \multicolumn{1}{c|}{\textbf{Memory Usage}} \\
         \hline\hline
         \#125 & 37.73KiB & 15min 40s & over 193GiB \\
         \hline
         \#147 & 5.13MiB & 15min 18s & over 193GiB \\
         \hline
         \#761 & 5.12MiB & 15min 45s & over 193GiB \\
         \hline
         \#148 & 5.12MiB & 15min 51s & over 193GiB \\
         \hline
         \#1732 & 29.52KiB & 39min 24s & 160GiB \\
         \hline
         \#1037 & 270.64KiB & 28min 53s & 143GiB \\
         \hline
         \#531 & 11.05KiB & 34min 04s & 142GiB \\
         \hline
         \#3281 & 164.60KiB & 28min 27s & 72GiB \\
         \hline
    \end{tabular}
    \end{adjustbox}
\end{table}

We set a memory-consumption threshold of 200 GB (190.73 GiB) for each
process and dynamically monitored memory usage at runtime. Once the total
memory consumption of a process and all of its child processes exceeded this
threshold, the process was terminated immediately. Accordingly, entries
reported as ``over 193 GiB'' in~\autoref{tab:costy_firmxray} reached this
limit. We also tested memory monitoring without this threshold and found that
some cases could exceed 400 GB of memory usage (\eg~case \#122).

\subsection{Rationales of Empirical Threshold Selections}
\label{app:threshold}

\paragraph{$\geq$10 valid functions for binary selection.} 
The threshold is used to distinguish firmware from non-code blobs. We observed that Ghidra occasionally identifies a few spurious functions in pure data files; requiring at least 10 valid functions effectively filters these false detections. Firmware with fewer than 10 valid functions is also unlikely to be meaningful for emulation or security analysis.
To verify our classification, we used the FirmLine~\cite{balgavy2024firmline} method, which employs cpu\_rec~\cite{granboulan2017cpurec} for ISA architecture detection, followed by Radare2~\cite{radare2} for verification. The results were consistent: the method identified 4,497 ARM Cortex-M binaries, all of which are included in our set of 4,571 ARM Cortex-M samples.

\paragraph{10 unique instructions in first 10,000 executed instructions for base address verification in stage 1.}
This serves as a preliminary check for candidate base addresses. Incorrect base addresses typically fail immediately (e.g., invalid memory accesses) or enter trivial loops, resulting in very few unique instructions. The final validation is performed in Stage 2, where unsuccessful firmware initialization further rejects incorrect configurations.

\paragraph{$\geq$10\% coverage for Stage-4.}
The 10\% threshold is a practical sampling criterion. Since Stage 4 (FirmRCA) is
computationally expensive, we prioritized higher-coverage crashes, which are
more likely to have progressed beyond initialization and are therefore more likely to be real
crashes. We agree that this introduces potential bias.

\subsection{Comparison of \fuzzware With and Without \gdma}
\label{app:gdma_comp}

In 35 test cases, \gdma identified DMA models, while results for other cases
matched the original \fuzzware results. Among the cases with recognized DMA
models, \gdma produced 920 additional crashes and increased coverage by only
1.09\% on average. Detailed comparisons are provided
in~\autoref{tab:gdma_comparison}.

\begin{table}[H]
    \centering
    \caption{Fuzzing results for \fuzzware with and without \gdma. Only the 27 cases with recognized DMA models and observable result differences are listed.}
    \label{tab:gdma_comparison}
    \begin{adjustbox}{width=0.9\columnwidth}
    \begin{tabular}{|r|r|r|r|r|}
         \hline
         \multicolumn{1}{|c|}{\multirow{2}{*}{\textbf{No.}}} & \multicolumn{2}{c|}{\textbf{Without \gdma}} & \multicolumn{2}{c|}{\textbf{With \gdma}} \\
         \cline{2-5}
         & \textbf{\#Crashes} & \textbf{Coverage} & \textbf{\#Crashes} & \textbf{Coverage} \\
         \hline\hline
         \#152 & 399 & 15.9\% & 649 & 15.9\% \\
         \hline
         \#1435 & 0 & 7.3\% & 0 & 7.8\% \\
         \hline
         \#1436 & 0 & 6.7\% & 0 & 7.1\% \\
         \hline
         \#1437 & 0 & 7.0\% & 0 & 7.5\% \\
         \hline
         \#1529 & 1,625 & 12.4\% & 1,651 & 12.7\% \\
         \hline
         \#1534 & 1,453 & 12.1\% & 2,164 & 12.8\% \\
         \hline
         \#1535 & 846 & 12.5\% & 1,971 & 12.9\% \\
         \hline
         \#1842 & 0 & 52.6\% & 0 & 55.3\% \\
         \hline
         \#1866 & 323 & 6.0\% & 456 & 6.1\% \\
         \hline
         \#2454 & 0 & 3.2\% & 0 & 4.3\% \\
         \hline
         \#2928 & 0 & 2.7\% & 0 & 3.7\% \\
         \hline
         \#2929 & 282 & 5.0\% & 462 & 5.4\% \\
         \hline
         \#3432 & 5,435 & 19.8\% & 140 & 18.2\% \\
         \hline
         \#3584 & 215 & 13.7\% & 888 & 15.2\% \\
         \hline
         \#3585 & 640 & 15.2\% & 884 & 15.3\% \\
         \hline
         \#3610 & 3,323 & 19.2\% & 1,958 & 18.7\% \\
         \hline
         \#3612 & 3,805 & 22.1\% & 3,247 & 22.3\% \\
         \hline
         \#6922 & 22 & 14.7\% & 395 & 15.6\% \\
         \hline
         \#7591 & 160 & 7.6\% & 234 & 12.9\% \\
         \hline
         \#7675 & 23 & 5.3\% & 5 & 5.3\% \\
         \hline
         \#9138 & 0 & 4.8\% & 0 & 4.8\% \\
         \hline
         \#9668 & 0 & 11.7\% & 0 & 11.8\% \\
         \hline
         \#11767 & 91 & 5.2\% & 243 & 6.1\% \\
         \hline
         \#13261 & 149 & 5.3\% & 278 & 6.1\% \\
         \hline
         \#17949 & 0 & 11.9\% & 0 & 14.0\% \\
         \hline
         \#18324 & 252 & 7.7\% & 256 & 13.1\% \\
         \hline
         \#18803 & 1,328 & 9.8\% & 5,410 & 16.0\% \\
         \hline
    \end{tabular}
    \end{adjustbox}
\end{table}

\subsection{Mapping Tool Exceptions to General Crash and Hang Types}
\label{app:crash/hang-map}

\subsubsection{\textbf{\fuzzware}}

\begin{itemize}
    \item \textbf{UR}: \texttt{UC\_ERR\_READ\_UNMAPPED}.
    \item \textbf{UW}: \texttt{UC\_ERR\_WRITE\_UNMAPPED}.
    \item \textbf{UF}: \texttt{UC\_ERR\_FETCH\_UNMAPPED}.
    \item \textbf{RV}: \texttt{UC\_ERR\_READ\_PROT}.
    \item \textbf{WV}: \texttt{UC\_ERR\_WRITE\_PROT}.
    \item \textbf{EV}: \texttt{UC\_ERR\_FETCH\_PROT}.
    \item \textbf{UI}: \texttt{UC\_ERR\_INSN\_INVALID}.
    \item \textbf{Hang}: No input is consumed within the first 3,000,000 executed basic blocks.
    \item \textbf{Other}: Other exceptions, such as \texttt{Unknown Crash} and \texttt{INCOMPLETE CONSUMPTION}.
\end{itemize}

\subsubsection{\textbf{\hoedur}}

\begin{itemize}
    \item \textbf{UR+UW+RV+(Partial)UF}: \texttt{Crash}, including unmapped fetches in the address range [0, 0x1000).
    \item \textbf{(Partial)UF+EV}: \texttt{NonExecutable}, excluding UF in \texttt{Crash}.
    \item \textbf{WV}: \texttt{RomWrite}.
    \item \textbf{Hang}: \texttt{Hang}, where the first 3,000,000 basic blocks from the start of emulation consume no input, or no input is consumed within 150,000 basic blocks after the most recent input consumption.
    \item \textbf{Other}: Other exceptions, including \texttt{RUNTIME\_ERROR}, \texttt{FAILED\_EXIT}, \texttt{InfiniteSleep}, and \texttt{Reset}. 
\end{itemize}

\subsubsection{\textbf{\multifuzz}}

\begin{itemize}
    \item \textbf{UR}: \texttt{ReadUnmapped}.
    \item \textbf{UW}: \texttt{WriteUnmapped}.
    \item \textbf{UF+EV}: \texttt{ExecViolation}.
    \item \textbf{RV}: \texttt{ReadPerm}.
    \item \textbf{WV}: \texttt{WritePerm}.
    \item \textbf{UI}: \texttt{Syscall} and \texttt{UnimplementedOp}.
    \item \textbf{Hang}: \texttt{Hang}, same as \hoedur, \texttt{INTERRUPT\_LIMIT}.
    \item \textbf{Other}: Other exceptions, including \texttt{InternalError}, \texttt{InvalidOpSize} (invalid instruction opcode size), \texttt{ExecUnaligned} (execution at an unaligned memory address), \texttt{UnknownError}, \texttt{InvalidInstruction}, and \texttt{SelfModifyingCode}.
\end{itemize}

\subsection{Crash Counts in Minor ``Other'' Categories}
\label{app:other-crash}
\begin{table}[H]
    \centering
    \caption{Crash counts in minor ``Other'' categories for \hoedur.}
    \label{tab:hoedur_minor}
    \begin{tabular}{|c|r|r|}
         \hline
         \textbf{Crash Type} & \textbf{\#Crashes} & \textbf{\#Samples} \\
         \hline\hline
         \textbf{Reset} & 15,203 & 103 \\
         \hline
         \textbf{Total} & 15,203 & 103 \\
         \hline
    \end{tabular}
\end{table}

The \texttt{Reset} category occurs when an unhandled hardware reset is
triggered by setting the \texttt{VECTRESET} field in the \texttt{AIRCR}
register, as in~\autoref{sec:gaps} (P12).

\begin{table}[H]
    \centering
    \caption{Crash counts in minor ``Other'' categories for \multifuzz.}
    \label{tab:multifuzz_minor}
    \begin{tabular}{|c|r|r|}
         \hline
         \textbf{Crash Type} & \textbf{\#Crashes} & \textbf{\#Samples} \\
         \hline\hline
         \textbf{ExecUnaligned} & 1 & 1 \\
         \hline
         \textbf{InvalidInstruction} & 1,114 & 35 \\
         \hline
         \textbf{InvalidOpSize} & 1 & 1 \\
         \hline
         \textbf{SelfModifyingCode} & 379 & 21 \\
         \hline
         \textbf{InternalError} & 61 & 13 \\
         \hline
         \textbf{Total} & 1,556 & 45 \\
         \hline
    \end{tabular}
\end{table}

The \texttt{ExecUnaligned}, \texttt{InvalidOpSize}, and
\texttt{SelfModifyingCode} categories are attributed to disassembly
implementation problems in \multifuzz. The \texttt{InvalidInstruction} error
arises from improper handling of mixed ARM/Thumb instructions, as discussed
in~\autoref{sec:gaps} (P12). The causes of \texttt{InternalError} remain
unidentified.

\subsection{Crash and Hang De-duplication Strategies}
\label{app:dedup_strategy}

\fuzzware does not implement crash deduplication. \hoedur and \multifuzz use
distinct deduplication strategies tailored to specific crash types. The
deduplication indicators used by \hoedur and \multifuzz are summarized
in~\autoref{tab:dedup_strategy_hoedur}
and~\autoref{tab:dedup_strategy_multifuzz}, respectively.

\begin{table}[H]
    \centering
    \caption{Crash and hang deduplication strategy used by \hoedur.}
    \label{tab:dedup_strategy_hoedur}
    \begin{tabular}{|c|c|}
         \hline
         \textbf{Crash Type} & \textbf{Deduplication Indicator} \\
         \hline\hline
         \textbf{UF}/\textbf{EV} & PC at crash site \\
         \hline
         \textbf{WV} & Write address leading to crash \\
         \hline
         \textbf{UR}/\textbf{UW}/\textbf{RV} & PC and LR at crash site \\
         \hline
    \end{tabular}
\end{table}

\begin{table}[H]
    \centering
    \caption{Crash and hang deduplication strategy used by \multifuzz.}
    \label{tab:dedup_strategy_multifuzz}
    \begin{tabular}{|c|c|}
         \hline
         \textbf{Crash Type} & \textbf{Deduplication Indicator} \\
         \hline\hline
         \textbf{Hang} & Caller PC of the hang function \\
         \hline
         \textbf{UF}/\textbf{EV} & PC of the last instruction executed before the crash \\
         \hline
         \textbf{UR}/\textbf{UW}/\textbf{RV}/\textbf{WV} & PC at crash site \\
         \hline
         \textbf{Other} & Tool-specific exit reason \\
         \hline
    \end{tabular}
\end{table}

\subsection{Detailed Fuzzing Initialization Results Across Seeds}
The same seed can cause different crash types across tools, and a single
sample may crash for different reasons under different seeds. Thus,
``Total'' denotes the number of unique samples.

\label{app:init-fail}

\begin{table}[H]
    \centering
    \caption{Seed-level initialization failures for \fuzzware.}
    \label{tab:fuzzware_seed_fail}
    \begin{adjustbox}{width=\columnwidth}
    \begin{tabular}{|r|r|r|r|r|r|r|r|r|r|r|r|}
         \hline
         \multicolumn{1}{|c|}{\textbf{Type}} & \multicolumn{1}{c|}{\textbf{UW}} & \multicolumn{1}{c|}{\textbf{UR}} & \multicolumn{1}{c|}{\textbf{UF}} & \multicolumn{1}{c|}{\textbf{RV}} & \multicolumn{1}{c|}{\textbf{WV}} & \multicolumn{1}{c|}{\textbf{EV}} & \multicolumn{1}{c|}{\textbf{UI}} & \multicolumn{1}{c|}{\textbf{Hang}} & \multicolumn{1}{c|}{\textbf{Other}} & \multicolumn{1}{c|}{\textbf{Total}} \\
         \hline\hline
         0 seed & 181 & 332 & 20 & 1 & 15 & 0 & 178 & 53 & 61 & 841 \\
         \hline
         1 seed & 183 & 367 & 48 & 1 & 15 & 0 & 189 & 53 & 97 & 953 \\
         \hline
         0/1 seed & 203 & 346 & 42 & 1 & 15 & 0 & 175 & 53 & 92 & 927 \\
         \hline
         Total & 208 & 393 & 49 & 1 & 15 & 0 & 194 & 53 & 96 & 997 \\
         \hline
    \end{tabular}
    \end{adjustbox}
\end{table}

\begin{table}[H]
    \centering
    \caption{Failure counts in minor ``Other'' categories for \fuzzware.}
    \label{tab:fuzzware_seed_minor}
    \begin{tabular}{|c|r|r|}
         \hline
         \textbf{Crash Type} & \textbf{\#Samples} \\
         \hline\hline
         \textbf{Unknown Crash} & 25 \\
         \hline
         \textbf{INCOMPLETE\_CONSUMPTION} & 70 \\
         \hline
         \textbf{UC\_ERR\_OK} & 9 \\
         \hline
         \textbf{AIRCR Write Error} & 1 \\
         \hline
         \textbf{Total} & 105 \\
         \hline
    \end{tabular}
\end{table}

\begin{table}[H]
    \centering
    \caption{Seed-level initialization failures for Fuzzware+\gdma.}
    \label{tab:gdma_seed_fail}
    \begin{adjustbox}{width=\columnwidth}
    \begin{tabular}{|r|r|r|r|r|r|r|r|r|r|r|r|}
         \hline
         \multicolumn{1}{|c|}{\textbf{Type}} & \multicolumn{1}{c|}{\textbf{UW}} & \multicolumn{1}{c|}{\textbf{UR}} & \multicolumn{1}{c|}{\textbf{UF}} & \multicolumn{1}{c|}{\textbf{RV}} & \multicolumn{1}{c|}{\textbf{WV}} & \multicolumn{1}{c|}{\textbf{EV}} & \multicolumn{1}{c|}{\textbf{UI}} & \multicolumn{1}{c|}{\textbf{Hang}} & \multicolumn{1}{c|}{\textbf{Other}} & \multicolumn{1}{c|}{\textbf{Total}} \\
         \hline\hline
         0 seed & 181 & 330 & 21 & 1 & 15 & 0 & 180 & 54 & 60 & 842 \\
         \hline
         1 seed & 183 & 361 & 52 & 1 & 15 & 0 & 195 & 54 & 97 & 958 \\
         \hline
         0/1 seed & 201 & 336 & 46 & 1 & 16 & 0 & 180 & 54 & 92 & 926 \\
         \hline
         Total & 209 & 388 & 54 & 1 & 16 & 0 & 199 & 54 & 97 & 1006 \\
         \hline
    \end{tabular}
    \end{adjustbox}
\end{table}

\begin{table}[H]
    \centering
    \caption{Failure counts in minor ``Other'' categories for Fuzzware+\gdma.}
    \label{tab:fuzzware_gdma_seed_minor}
    \begin{tabular}{|c|r|r|}
         \hline
         \textbf{Crash Type} & \textbf{\#Samples} \\
         \hline\hline
         \textbf{Unknown Crash} & 25 \\
         \hline
         \textbf{INCOMPLETE\_CONSUMPTION} & 71 \\
         \hline
         \textbf{UC\_ERR\_OK} & 9 \\
         \hline
         \textbf{Physical Time Exceeded} & 5 \\
         \hline
         \textbf{Total} & 110 \\
          \hline
    \end{tabular}
\end{table}

The \texttt{Unknown Crash} category results from memory overlap, as
discussed in~\autoref{sec:gaps} (P5). \texttt{INCOMPLETE\_CONSUMPTION} is
caused by an unhandled hardware reset, as noted in~\autoref{sec:gaps} (P12). \texttt{AIRCR Write Error} occurs when firmware
writes an invalid value to the AIRCR register; according to the ARM
architecture, the high 16 bits of AIRCR must be \texttt{0x5fa}. The cause of
\texttt{UC\_ERR\_OK} remains unidentified.

\begin{table}[H]
    \centering
    \caption{Seed-level initialization failures for \hoedur.}
    \label{tab:hoedur_seed_fail}
    \begin{adjustbox}{width=\columnwidth}
    \begin{tabular}{|r|r|r|r|r|r|r|r|r|}
         \hline
         \multicolumn{1}{|c|}{\textbf{Type}} & \multicolumn{1}{c|}{\textbf{UR+UW+RV}} & \multicolumn{1}{c|}{\textbf{WV}} & \multicolumn{1}{c|}{\textbf{UF+EV}} & \multicolumn{1}{c|}{\textbf{UI}} & \multicolumn{1}{c|}{\textbf{Hang}} & \multicolumn{1}{c|}{\textbf{Other}} & \multicolumn{1}{c|}{\textbf{Total}} \\
         \hline\hline
         0 seed     & 596 & 15 & 5 & 0 & 61 & 51 & 728 \\
         \hline
         1 seed     & 702 & 13 & 4 & 0 & 61 & 84 & 864 \\
         \hline
         0/1 seed   & 763 & 15 & 4 & 0 & 65 & 82 & 929 \\
         \hline
         Total      & 805 & 17 & 5 & 0 & 65 & 102 & 983 \\
         \hline
    \end{tabular}
    \end{adjustbox}
\end{table}

\begin{table}[H]
    \centering
    \caption{Failure counts in minor ``Other'' categories for \hoedur.}
    \label{tab:hoedur_seed_minor}
    \begin{tabular}{|c|r|r|}
         \hline
         \textbf{Crash Type} & \textbf{\#Samples} \\
         \hline\hline
         \textbf{Reset} & 76 \\
         \hline
         \textbf{FAILED\_EXIT\_134} & 25 \\
         \hline
         \textbf{RUNTIME\_ERROR} & 1 \\
          \hline
         \textbf{Total} & 102 \\
         \hline
    \end{tabular}
\end{table}

The \texttt{FAILED\_EXIT\_134} category results from memory overlap, as
discussed in~\autoref{sec:gaps} (P5). \texttt{Reset} is caused by an
unhandled hardware reset, as noted in~\autoref{sec:gaps} (P12). The
\texttt{RUNTIME\_ERROR} case is caused by a configuration-file parsing error.

\begin{table}[H]
    \centering
    \caption{Seed-level initialization failures for \multifuzz.}
    \label{tab:multifuzz_seed_fail}
    \begin{adjustbox}{width=\columnwidth}
    \begin{tabular}{|r|r|r|r|r|r|r|r|r|r|}
         \hline
         \multicolumn{1}{|c|}{\textbf{Type}} & \multicolumn{1}{c|}{\textbf{UW}} & \multicolumn{1}{c|}{\textbf{UR}} & \multicolumn{1}{c|}{\textbf{UF+EV}} & \multicolumn{1}{c|}{\textbf{RV}} & \multicolumn{1}{c|}{\textbf{WV}} & \multicolumn{1}{c|}{\textbf{UI}} & \multicolumn{1}{c|}{\textbf{Hang}} & \multicolumn{1}{c|}{\textbf{Other}} & \multicolumn{1}{c|}{\textbf{Total}} \\
         \hline\hline
         0 seed     & 203 & 342 & 18 & 1 & 14 & 16 & 62 & 67 & 723 \\
         \hline
         1 seed     & 207 & 406 & 44 & 1 & 15 & 17 & 62 & 100 & 852 \\
         \hline
         0/1 seed   & 240 & 452 & 34 & 1 & 17 & 17 & 62 & 105 & 928 \\
         \hline
         Total      & 246 & 483 & 46 & 1 & 17 & 17 & 62 & 123 & 969 \\
         \hline
    \end{tabular}
    \end{adjustbox}
\end{table}

\begin{table}[H]
    \centering
    \caption{Failure counts in minor ``Other'' categories for \multifuzz.}
    \label{tab:multifuzz_seed_minor}
    \begin{tabular}{|c|r|r|}
         \hline
         \textbf{Crash Type} & \textbf{\#Samples} \\
         \hline\hline
         \textbf{SelfModifyingCode} & 16 \\
         \hline
         \textbf{InternalError} & 4 \\
         \hline
         \textbf{UnknownError} & 1 \\
         \hline
         \textbf{No seed results} & 25 \\
         \hline
         \textbf{INCOMPLETE\_CONSUMPTION} & 77 \\
         \hline
         \textbf{Total} & 123 \\
         \hline
    \end{tabular}
\end{table}

The \texttt{No seed results} category arises from memory overlap, as
explained in~\autoref{sec:gaps} (P5). The
\texttt{INCOMPLETE\_CONSUMPTION} error results from an unhandled hardware
reset, as noted in~\autoref{sec:gaps} (P12). \texttt{SelfModifyingCode} is
attributed to disassembly issues. The causes of \texttt{InternalError} and
\texttt{UnknownError} remain unidentified.

\begin{table}[H]
    \centering
    \caption{Seed-level initialization failures for \aid.}
    \label{tab:aidfuzzer_seed_fail}
    \begin{adjustbox}{width=0.8\columnwidth}
    \begin{tabular}{|c|c|c|c|c|c|}
         \hline
         \multicolumn{1}{|c|}{\textbf{Type}} & \multicolumn{1}{c|}{\textbf{Crash}} & \multicolumn{1}{c|}{\textbf{Hang}} & \multicolumn{1}{c|}{\textbf{Model Failure}} & \multicolumn{1}{c|}{\textbf{Other}} & \multicolumn{1}{c|}{\textbf{Total}} \\
         \hline\hline
         0 seed     & 1,546 & 769 & 120 & 25 & 2,460 \\
         \hline
         1 seed     & 1,546 & 769 & 119 & 25 & 2,459 \\
         \hline
         0/1 seed   & 1,546 & 769 & 119 & 25 & 2,459 \\
         \hline
         Total      & 1,546 & 769 & 120 & 25 & 2,460 \\
         \hline
    \end{tabular}
    \end{adjustbox}
\end{table}

\aid crashes primarily stem from incomplete memory mapping of the system
peripheral region, as noted in~\autoref{sec:gaps} (P5). Hangs and model
failures are mainly attributed to \aid's IRQ modeling issues. 

\subsection{Heuristic Attribution of False Crashes and Hangs}
\label{app:problem-attribution}

We use the following heuristics to estimate the false crashes and hangs
associated with P4--P9 and P12 during fuzzing. These criteria capture
highly suspicious signatures observed when the corresponding problems occur;
they are not exact causal classifiers. Therefore, a matching signature does not guarantee that the problem
is the true cause, while failures without the signature may still result from
it. Accordingly, some problems cannot be quantified reliably, and the reported
problem ratios may not sum to 100\% in Appendix~\ref{app:crash-hang}. We use these heuristics only to approximate
the prevalence of the identified problems.

\paragraph{P4: Unsupported ISA Instructions.}
Crashes resulting from Unsupported-instruction (\textbf{UI}) can be directly identified through tool-specific exception types. For instance, \fuzzware indicates this with \texttt{UnimplementedOp}, whereas \multifuzz reports either \texttt{Syscall} or \texttt{UnimplementedOp} exceptions.

\paragraph{P5: Inaccurate Memory Layout.}
We associate a crash with P5 when it accesses an address that is unmapped by
the tool but falls within a valid ARM memory region that may legitimately be
used by firmware. We consider the following regions:
(\texttt{0x10000000-0x1FFFFFFF}), external-bus regions
(\texttt{0x60000000-0xDFFFFFFF}), and Code/Flash regions
\texttt{0x08000000-0x0FFFFFFF} and
\texttt{0x00000100-0x07FFFFFF}, excluding ranges already occupied by the
loaded firmware image.

\paragraph{P6: Absolute Speed Gap.}
P6 manifests primarily as hangs. However, hangs may also arise from inaccurate
peripheral responses (P8) or incomplete peripheral semantics (P9). Since these
causes cannot be reliably distinguished from the hang outcome alone, we do not
estimate the proportion of hangs attributable specifically to P6.

\paragraph{P7: Incorrect Interrupt Timing.}
We associate a false crash with P7 when the crash occurs in interrupt context
and involves an uninitialized-pointer access. Interrupt context is identified
using tool-specific runtime state, such as
\texttt{active\_irq > 0} in \multifuzz or
\texttt{err\_in\_interrupt > 0} in GDMA-enabled executions. We
consider the resulting invalid-read (\textbf{RV}) failures caused by
uninitialized pointers.

\paragraph{P8/P9: Inaccurate Peripheral Responses and Incomplete Peripheral Modeling.}
For false crashes,
we use the crash address as a heuristic indicator. We associate crashes with
P8/P9 when the faulting memory access or instruction fetch targets
(i) the peripheral region (\texttt{0x40000000-0x5FFFFFFF}),
(ii) the private/system-peripheral region
(\texttt{0xE0000000-0xFFFFFFFF}), or
(iii) an otherwise unmapped region
(\texttt{0x20100000-0x3FFFFFFF}).
The last region is included because inaccurate peripheral values may themselves
be interpreted as addresses or address-related metadata, as in DMA descriptors
or the pointer-like peripheral-register semantics described in P9, causing the
eventual fault to occur outside the peripheral address space.
False hangs caused by P8 and P9 cannot be reliably distinguished from other
hang causes and are therefore not quantitatively attributed. 

\paragraph{P12: Implementation Limitations.}
Implementation-specific failures are identified using tool-specific exception
signatures corresponding to known implementation defects. For example,
\multifuzz failures caused by unsupported or incorrect ARM/Thumb mode
transitions appear as \texttt{ReadPerm} exceptions with the
\texttt{UnknownError::READ\_PROT} subtype.

\subsection{Details of false-crash counts and problem categories across evaluated tools in stage 3.}

\label{app:crash-hang}

\paragraph{Crash-type legend.} \textit{Unmapped Access} denotes a read (\textbf{UR}), write (\textbf{UW}), or instruction fetch (\textbf{UF}) from an address outside all memory regions defined in~\autoref{tab:memorymap}; 
\textit{Permission Violation} denotes a read (\textbf{RV}), write (\textbf{WV}), or execution (\textbf{EV}) at an address lacking the corresponding permission under the configured memory-access rules;
\textit{Unsupported Instruction} (\textbf{UI}) denotes execution of an instruction unsupported by the emulator.
Mappings from tool-specific crash types to these categories are provided in Appendix~\ref{app:crash/hang-map}.
Breakdowns of the other crash types are provided in Appendix~\ref{app:other-crash}.

\begin{table}[H]
\centering
\caption{False-crash counts and root-cause categories for GDMA (Fuzzware).}
\label{tab:fp-type-fuzzware}
\begin{adjustbox}{width=\columnwidth}
\begin{tabular}{|c|r|r|l|}
    \hline
    \textbf{Crash Type} &
    \textbf{\#Crashes} &
    \textbf{\#Samples} &
    \textbf{Problems (Ratio\%)} \\
    \hline\hline

    \textbf{UR} & 152,680 & 374
    & P5 (82\%) P8/9 (4\%) \\
    \hline
    \textbf{UW} & 22,767 & 101
    & P5 (84\%) P8/9 (16\%) \\
    \hline
    \textbf{RV} & 61 & 7
    & P8/9 (100\%) \\
    \hline
    \textbf{UF} & 14,673 & 68
    & P5 (14\%) P8/9 (9\%) \\
    \hline
    \textbf{EV} & 2,677 & 40
    & P5 (68\%) P8/9 (4\%) \\
    \hline
    \textbf{WV} & 6,602 & 31
    & P7 (85\%) P8/9 (13\%) P5 (2\%) \\
    \hline
    \textbf{UI} & 12,315 & 74
    & P4 (100\%) \\
    \hline
    \textbf{Total} & \textbf{211,775} & \textbf{521} & \\
    \hline
\end{tabular}
\end{adjustbox}
\end{table}

\begin{table}[H]
\centering
\caption{False-crash counts and root-cause categories for \multifuzz.}
\label{tab:fp-type-multifuzz}
\begin{adjustbox}{width=\columnwidth}
\begin{tabular}{|c|r|r|l|}
    \hline
    \textbf{Crash Type} &
    \textbf{\#Crashes} &
    \textbf{\#Samples} &
    \textbf{Problems (Ratio\%)} \\
    \hline\hline

    \textbf{UR} & 1,565 & 511
    & P5 (51\%) P7 (33\%) P8/9 (10\%) \\
    \hline
    \textbf{UW} & 680 & 125
    & P7 (42\%) P5 (35\%) P8/9 (15\%) \\
    \hline
    \textbf{RV} & 2,861 & 120
    & P12 (1\%) \\
    \hline
    \multirow{2}{*}{\textbf{UF + EV}} & \multirow{2}{*}{57,465} & \multirow{2}{*}{177} & \multirow{2}{*}{P5 (64\%) P8/9 (32\%) P7 (2\%)} \\
     & & & \\
    \hline
    \textbf{WV} & 344 & 42
    & P7 (64\%) P5 (16\%) P8/9 (10\%) \\
    \hline
    \textbf{UI} & 1,549 & 64
    & P4 (100\%) \\
    \hline
    \textbf{Hang} & 2,987 & 1,301 & P6 \\
    \hline
    \textbf{Others} & 442 & 30
    & P7 (77\%) P5 (20\%) \\
    \hline
    \textbf{Total} & \textbf{67,893} & \textbf{1,488} & \\
    \hline
\end{tabular}
\end{adjustbox}
\end{table}

\begin{table}[H]
\centering
\caption{False-crash counts and root-cause categories for \hoedur.}
\label{tab:fp-type-hoedur}
\begin{adjustbox}{width=\columnwidth}
\begin{tabular}{|c|r|r|l|}
    \hline
    \textbf{Crash Type} &
    \textbf{\#Crashes} &
    \textbf{\#Samples} &
    \textbf{Problems (Ratio\%)} \\
    \hline\hline

    \textbf{UR+UW+RV+UF+EV} & 14,749 & 590
    & P5 (36\%) \\
    \hline
    \textbf{WV} & 471 & 29
    & P5 (92\%) \\
    \hline
    \textbf{Hang} & 27,987 & 1,065 & P6 \\
    \hline
    \textbf{Others} & 15,203 & 103 & - \\
    \hline
    \textbf{Total} & \textbf{58,410} & \textbf{1,219} & \\
    \hline
\end{tabular}
\end{adjustbox}
\end{table}

\end{document}